\documentclass[%
 aip,
 amsmath,amssymb,
 reprint,
]{revtex4-1}

\pdfoutput=1
\usepackage[usenames,dvipsnames]{xcolor}
\usepackage{hyperref}
\usepackage{soul}
\usepackage{tablefootnote}
\usepackage{mathtools}
\usepackage{graphicx}
\usepackage[english]{babel}
\usepackage{amsmath}
\usepackage{amssymb}
\usepackage{epstopdf}
\usepackage{diagbox}
\usepackage{bbm}
\usepackage{dsfont}
\usepackage{bbold}
\usepackage{color,soul}
\usepackage{dcolumn}
\usepackage{latexsym}
\usepackage{bm}
\usepackage{upgreek}
\usepackage{ulem}
\usepackage{listings}
\usepackage{cases}
\usepackage{enumerate}
\usepackage[version=3]{mhchem}
\usepackage{float}
\usepackage{csquotes}
\usepackage[caption=false]{subfig}
\definecolor{lg}{gray}{0.93}
\usepackage{xr-hyper}
\DeclareRobustCommand*{\citen}[1]{%
  \begingroup
    \romannumeral-`\x
    \setcitestyle{numbers}%
    \cite{#1}%
  \endgroup
}

\newcommand{\figplaceholder}[2][0.95\columnwidth]{%
  \IfFileExists{#2}{\includegraphics[width=#1]{#2}}{%
    \fbox{\parbox[c][4.5cm][c]{0.88\columnwidth}{\centering Missing figure file\\\texttt{#2}}}%
  }%
}

\begin{document}
\normalem


\title[]{Concentration-Dependent Speciation in \ce{ZnCl2} Electrolyte Solutions: Transferability of Machine-Learning Interatomic Potentials and Speciation Sensitivity of EXAFS}

\author{Junhan Chen}
\email{junhan.chen@pnnl.gov}
\affiliation
{\mbox{Integrated Discovery Sciences Directorate, Pacific Northwest National Laboratory, Richland, WA 99352, USA}}
\author{Luke D. Gibson}
\affiliation
{\mbox{Integrated Discovery Sciences Directorate, Pacific Northwest National Laboratory, Richland, WA 99352, USA}}
\author{Nadia N. Intan}
\affiliation
{\mbox{Integrated Discovery Sciences Directorate, Pacific Northwest National Laboratory, Richland, WA 99352, USA}}
\author{Darren M Driscoll}
\affiliation
{\mbox{X-ray Science Division, Argonne National Laboratory, Argonne, IL 60439, USA}}
\author{Mahalingam Balasubramanian}
\affiliation
{\mbox{X-ray Science Division, Argonne National Laboratory, Argonne, IL 60439, USA}}
\author{John L. Fulton}
\affiliation
{\mbox{Integrated Discovery Sciences Directorate, Pacific Northwest National Laboratory, Richland, WA 99352, USA}}
\author{Gregory K. Schenter}
\affiliation
{\mbox{Integrated Discovery Sciences Directorate, Pacific Northwest National Laboratory, Richland, WA 99352, USA}}
\author{Christopher J. Mundy}
\affiliation
{\mbox{Integrated Discovery Sciences Directorate, Pacific Northwest National Laboratory, Richland, WA 99352, USA}}
\author{Mohammadhasan Dinpajooh}
\email{hadi.dinpajooh@pnnl.gov}
\affiliation{\mbox{Integrated Discovery Sciences Directorate, Pacific Northwest National Laboratory, Richland, WA 99352, USA}}



\begin{abstract}
Predicting speciation in concentrated electrolytes requires accurate ion interactions and sufficient sampling of slowly interconverting coordination environments. We combine Zn K-edge extended X-ray absorption fine structure (EXAFS) measurements with long-timescale molecular dynamics (MD) driven by machine-learning interatomic potentials (MLIPs) to investigate aqueous \ce{ZnCl2} from dilute to 30~m. Using active learning, we train revPBE-based DeepMD potentials with D3 corrections included or excluded for Zn-containing pairs and compare them with DPMD-SCAN, classical MD (CMD), and MACE-MATPES-r2SCAN. We benchmark related EXAFS calculations against hydrated and chloride-coordinated Zn reference complexes, calculate spectra from MD ensembles, and examine Zn--Cl coordination free energies. Increasing concentration reduces octahedral Zn coordination, promotes replacement of water by chloride, and increases chloride-bridged association. Although all three DeepMD models yield similar EXAFS at 3~m, DPMD-SCAN shows greater transferability across concentrations, reproducing measured spectral trends, capturing key features of mixed \ce{ZnCl2}/LiCl spectra, and predicting octahedral fractions consistent with other studies. DFT treatment strongly affects coordination free energies and aggregation, while excluding Zn-related D3 corrections does not systematically improve EXAFS agreement. CMD shows larger spectral discrepancies and less consistent coordination trends. MACE-MATPES-r2SCAN improves upon CMD and produces spectra comparable to DPMD-SCAN, but DPMD-SCAN provides more consistent agreement and greater computational efficiency. Neighbor-resolved analysis reveals that distance disorder and interference among scattering paths can produce similar EXAFS signatures for distinct coordination ensembles. This highlights the importance of rigorous molecular sampling and ensemble-averaged spectral calculations to complement EXAFS when interpreting isolated versus chloride-bridged Zn speciation.
\end{abstract}

\maketitle

\section{Introduction}
\label{sec:intro}

Electrolyte speciation, defined as the distribution of ions among distinct coordination and association environments, plays a key role in determining the structural, thermodynamic, and transport properties of concentrated aqueous solutions.\cite{aurbach2000prototype,Gebauer-2008,Peng-2018,Finney-2019,Zhang-2020,baskin2020ion,Stefan-2024,Cao-2025,Stevenson-2026}
Predicting these populations across a broad concentration range requires an accurate description of competing ion–water and ion–ion interactions,\cite{Dill-2009,Vegt-2016,Paesani-2017,Pavel-2018,Baer-2016,Tim-2017,galib2017revisiting,duignan2020quantifying,Li-2025,Paesani-2026} which govern the relative free energies of different coordination states,\cite{Roy-2016,Wang-2020,Dinpajooh-2026-hno3} together with sufficient configurational sampling to obtain converged speciation statistics. These challenges are particularly pronounced for multivalent ions, where small errors in the description of these interactions can substantially alter the predicted properties.\cite{Netz-2010,Netz-2013,saxena2015multisite,Shuwen-2021,Panagiotopoulos-2023,Dinpajooh-2024,Dinpajooh-2024-2}

Aqueous \ce{ZnCl2} provides a representative system for investigating these challenges. At low chloride activity, \ce{Zn^{2+}} is predominantly hydrated in an approximately octahedral environment.\cite{Irish-1963,Dreier1986ZnEXAFS,DAngelo2011ZnCl2XAS,Dhakal2023ZnHalides} As concentration and chloride activity increase, chloride progressively replaces water in the first coordination shell, producing a distribution of \ce{[ZnCl_n(H2O)_m]^{2-n}} coordination environments.
At higher concentrations, these environments can become connected through shared chloride ligands, forming bridged and multinuclear Zn--Cl structures.\cite{Cao-2025} Consequently, the number of chloride neighbors around a Zn center does not necessarily identify a distinct monomeric species. Distinguishing local coordination from molecular connectivity is therefore important for understanding speciation in concentrated electrolytes.

Extended X-ray absorption fine structure (EXAFS) spectroscopy provides element-specific information about the local coordination environment of ions such as \ce{Zn^{2+}}, including the identities of neighboring atoms, interatomic distances, average coordination numbers, and structural disorder.\cite{Allen-1997,Allen-2000,Rehr2000XAFS,Spangberg2000MD_EXAFS,Newville2014XAFS}
However, extracting a unique speciation distribution from EXAFS is challenging because the measured spectrum represents an average over different coordination environments. Coordination numbers and Debye--Waller factors are correlated in conventional EXAFS fitting, and the relatively close ion--O and ion--Cl first-shell distances complicate their separation in real space.\cite{Dreier1986ZnEXAFS,DAngelo2011ZnCl2XAS}
Furthermore, isolated complexes and chloride-bridged multinuclear structures may have similar first-shell environments and therefore produce similar EXAFS signals. Thus, EXAFS provides important experimental constraints on local coordination but generally requires additional structural information to determine detailed molecular speciation.

Molecular simulations provide complementary information by generating structural ensembles from which coordination populations, molecular connectivity, and theoretical EXAFS spectra can be obtained.\cite{Spangberg2000MD_EXAFS,Wan2014MgXAS}
Ab initio molecular dynamics (AIMD) describes ion association and ligand exchange without imposing fixed bonding topologies, but its computational cost limits the accessible timescales and system sizes.\cite{Eric-2008-UO2,cauet2010structure}
Although AIMD is well suited to studying the hydration of individual ions, determining speciation in concentrated solutions requires extensive sampling of collective ionic configurations to obtain statistically converged populations.
Classical molecular dynamics (CMD) enables substantially longer simulations, but conventional fixed-charge force fields may not accurately capture the competing metal--water, metal--chloride, and solvent-mediated interactions.\cite{McDaniel2018Polarization,McDaniel2018IonCorrelation} Such inaccuracies can affect both the relative free energies of coordination states and the barriers separating them, potentially overestimating ligand-exchange barriers relative to first-principles predictions.\cite{Timko-2010,Lemkul-2016,Pavel-2018} Consequently, even long CMD trajectories may yield biased or poorly converged speciation populations.

Machine-learning interatomic potentials (MLIPs) offer a promising route to bridge the gap between AIMD and conventional classical force fields by approximating electronic-structure potential-energy surfaces at a fraction of the computational cost.\cite{Sun2016,Zhang2018DeepPotential,Wang2018DeePMDKit,MACE,batatia-2024,Gale-2025,Orb-2025,Cheng2024CACE,Cheng2025LatentEwald,Maxson-2024,Maxson-2024-2,Allgero-2023}
This approach enables simulations of larger electrolyte systems over longer timescales while retaining a first-principles-based description of the competing interactions that govern coordination equilibria.
For electrolyte solutions, an important question is whether the trained MLIP remains predictive across concentrations as local coordination and ion association environments change.
Here, transferability across concentrations refers to this ability to describe different concentration regimes without retraining at each composition, rather than transferability to unrelated chemical systems.\cite{batatia-2024,Orb-2025,Allgero-2023}
Such transferability is not guaranteed, because changes in concentration can introduce local environments that are inadequately represented in the training data.\cite{Hanseroth2025,Dinpajooh-2026-hno3}

Recent work on aqueous solutions demonstrated that the choice of training concentration can strongly influence transferability across concentrations,\cite{Hanseroth2025,Dinpajooh-2026-hno3} while studies of ionic-liquid mixtures have demonstrated related transferability across mixture compositions.\cite{Goodwin2024} In this regard, recent DeepMD simulations of aqueous \ce{ZnCl2} have extended to tens or hundreds of nanoseconds across a broad concentration range, revealing concentration-dependent changes in local coordination and ion association.\cite{Cao-2025}
Nevertheless, long trajectories alone do not guarantee converged speciation populations, particularly when transitions between coordination states are infrequent.
Simulations starting from independent initial configurations can help assess convergence, while targeted enhanced sampling, such as umbrella sampling along Zn--Cl separation and coordination number, can provide complementary information about association free energies and barriers.

The reliability of MLIP-based speciation predictions also depends on the underlying DFT description. Even when an MLIP accurately reproduces reference DFT energies and forces, errors in the exchange--correlation functional or dispersion treatment can alter the relative stability of competing coordination environments. Kostal \textit{et al.} demonstrated that conventional D3 dispersion corrections can distort the hydration structures of alkali and alkaline-earth cations because their dispersion coefficients are based on neutral-atom reference states.\cite{kostal2023common} Although \ce{Zn^{2+}} was not examined, these findings motivate assessing the influence of dispersion corrections on Zn coordination environments. These considerations highlight the importance of evaluating coordination free energies, the convergence of speciation populations, and agreement with experimental structural measurements, rather than relying solely on the accuracy of reference energies and forces.

Recent advances in computational spectroscopy have strengthened the connection between simulated structures and X-ray absorption measurements.\cite{Chen2021MLScattering,Chen2024RobustXAS} For aqueous \ce{ZnCl2}, Cao \textit{et al.} combined experimental Zn K-edge X-ray absorption near-edge structure (XANES) spectroscopy with an interpretable graph-neural-network model trained on first-principles spectra to relate local coordination environments to near-edge spectral features.\cite{Cao2026ZnCl2XAS} A complementary challenge is to determine how sensitively experimental extended X-ray absorption fine structure (EXAFS) spectra constrain the structural ensembles and coordination populations predicted by different interaction models. Because EXAFS reflects an ensemble average over coexisting local environments, independent structural analyses are essential for assessing the extent to which spectral agreement constrains the underlying speciation.

In this work, we combine active learning,\cite{ARCANN,Schran-2020} long-timescale molecular dynamics, targeted enhanced sampling, and experimental EXAFS spectroscopy to investigate the concentration-dependent structure and speciation of aqueous \ce{ZnCl2}. We examine how the choice of DFT functional and dispersion treatment influences the predictions of DeepMD-based potentials and compare their accuracy and performance with a universal MLIP and conventional classical MD. Independent simulations and coordination free-energy calculations provide complementary assessments of sampling convergence and the stability of competing coordination environments. We focus on comparing ensemble-averaged EXAFS spectra calculated from MD configurations with experimental measurements across a broad concentration range, including highly concentrated solutions where structural interpretation becomes increasingly challenging. Concentration-dependent changes are further examined using radial distribution functions, coordination numbers, and local coordination geometry. Section~II describes the computational and experimental methods, Sec.~III presents the results and discussion, and Sec.~IV summarizes the main conclusions.

\section{Numerical \& Experimental Details}
\label{numerical_detail}

\subsection{Custom-trained ML Training Methodology}
\label{ML_detail}

We employed an iterative active-learning strategy using the ArcaNN software package\cite{ARCANN} to explore the complex free-energy landscape of aqueous \ce{ZnCl2} solutions and train the machine-learning interatomic potential (MLIP).
The initial training configurations were obtained from enhanced-sampling {\it ab initio} molecular dynamics (AIMD) simulations of aqueous \ce{ZnCl2} performed in this work.
The Zn--Cl separation and the chloride coordination number around \ce{Zn^{2+}} were used as collective variables, as described in Section~\ref{Sec:AIMD}.
We also used configurations from a previously reported neural-network-potential study of aqueous \ce{ZnCl2},\cite{Cao-2025} which provided a broad set of equilibrium solution structures across various concentrations.
These initial configurations served only as the seed dataset.
The subsequent active-learning cycles systematically expanded the training set by identifying and labeling poorly represented regions of configuration space, including subtle changes in local coordination and charge-state environments, as described below.
The compositions of the aqueous \ce{ZnCl2} systems included in the training dataset are summarized in Table~\ref{tab:training_compositions}.

For labeling all configurations, single-point calculations of the atomic forces and total energies were performed using the CP2K QuickStep module\cite{CP2K-2020} within density functional theory (DFT).\cite{Hohenberg-Kohn-1964,Khon-Sham-1965}
Various exchange-correlation functionals and dispersion treatments were considered.
The functionals included revPBE\cite{zhang1998comment}.
Motivated by the recent work of Jungwirth {\it et al.},\cite{Kalvoda2024Hydration,kostal2023common} D3 dispersion contributions\cite{grimme2010consistent} involving the \ce{Zn^{2+}} cation were excluded from the main revPBE calculations.
Valence electrons were treated explicitly, while the core electrons of all atoms were represented by norm-conserving Goedecker--Teter--Hutter (GTH) pseudopotentials.\cite{GTH1}
Wavefunctions were expanded using a mixed Gaussian-and-plane-wave scheme with TZV2P-MOLOPT basis sets\cite{molopt} for all elements and a 600 Ry auxiliary plane-wave cutoff for the electron density.\cite{Chris-2011,Chris-2011-2}
Self-consistent-field convergence was accelerated using the orbital-transformation (OT)\cite{VandeVondele-2003} and Direct Inversion in the Iterative Subspace (DIIS) algorithms,\cite{Pulay-1980,Pulay-1982} with $\epsilon_{\rm SCF}=10^{-6}$.
Single-point calculations were carried out in periodic orthorhombic cells whose dimensions matched those of the molecular configurations sampled from the earlier trajectories (see Table~\ref{tab:training_compositions}).
The \texttt{CELL\_REF} keyword in CP2K was used to define a fixed reference cell with a box length of 25~\AA, ensuring consistency in the reference data used for training the MLIP, especially when cell fluctuations were present in NpT simulations.
The resulting high-fidelity energy and force datasets served as the ground-truth labels for training the MLIP.

\begin{table}[t]
\centering
\caption{Compositions of the aqueous \ce{ZnCl2} training systems.
The composition column lists
$N_{\rm Zn}/N_{\rm Cl}/N_{\rm O}/N_{\rm H}$.
For nonstoichiometric water configurations, the nominal molality $m$ is calculated using $N_{\rm O}$ water equivalents.}
\label{tab:training_compositions}
\scriptsize
\setlength{\tabcolsep}{2.5pt}
\renewcommand{\arraystretch}{1.05}
\begin{tabular}{@{}ccccc@{}}
\hline
System &
$N_{\rm Zn}/N_{\rm Cl}/N_{\rm O}/N_{\rm H}$ &
$l$ [\AA] &
$c$ [M] &
$m$ [mol\,kg$^{-1}$] \\
\hline
\multicolumn{5}{c}{Stoichiometric water configurations} \\
\hline
S1 & $1/2/93/186$   & 14.3000 & 0.568  & 0.597  \\
S2 & $1/2/83/166$   & 13.3603 & 0.696  & 0.669  \\
S3 & $1/2/53/106$   & 11.4980 & 1.092  & 1.047  \\
S4 & $3/6/48/96$    & 11.6086 & 3.184  & 3.469  \\
S5 & $6/12/51/102$  & 12.1259 & 5.588  & 6.530  \\
S6 & $10/20/48/96$  & 12.8629 & 7.802  & 11.564 \\
S7 & $12/24/36/72$  & 12.5973 & 9.968  & 18.503 \\
S8 & $13/26/24/48$  & 12.1056 & 12.168 & 30.067 \\
\hline
\multicolumn{5}{c}{Proton-rich and proton-deficient configurations} \\
\hline
S9--S10 & $1/4/264/530$ & 20.0000 & 0.208 & 0.210 \\
S11     & $1/3/92/185$  & 14.3000 & 0.568 & 0.603 \\
S12     & $1/1/265/529$ & 20.0000 & 0.208 & 0.209 \\
S13     & $1/0/95/188$  & 14.3000 & 0.568 & 0.584 \\
S14     & $1/1/94/187$  & 14.3000 & 0.568 & 0.591 \\
S15     & $1/3/92/185$  & 14.3000 & 0.568 & 0.603 \\
S16     & $1/4/91/184$  & 14.3000 & 0.568 & 0.610 \\
S17     & $1/5/90/183$  & 14.3000 & 0.568 & 0.617 \\
\hline
\end{tabular}
\end{table}

The active-learning configurations were generated using the ArcaNN workflow\cite{ARCANN} coupled to DeepMD-kit.\cite{DeepMD-2018,DeepMD-2023}
We targeted the relevant free-energy landscapes using collective variables implemented in PLUMED.\cite{PLUMED}
The collective variables were chosen to describe Zn--Cl separation, Zn--Cl coordination, and Zn--water coordination, together with configurations representing pH-dependent speciation.
ArcaNN was used to organize the iterative sampling and labeling procedure, while the active-learning cycle itself was carried out using DeepMD models.
Each active-learning iteration consisted of three steps: training, exploration, and labeling.
In the training step, three DeepMD models (see below) were trained on the same reference dataset but with different random seeds for parameter initialization.
This ensemble was then used during exploration to estimate model uncertainty from the spread in the predicted atomic forces.
Configurations with intermediate model deviation ($0.2 < \sigma < 0.7$) were selected for labeling by single-point DFT calculations using CP2K\cite{CP2K-2020} and were added to the training dataset.
Configurations with very large model deviations were excluded from labeling because they typically corresponded to poorly sampled or unphysical high-energy structures.
This procedure was repeated until the number of newly selected configurations became small and reached a plateau, and the sampled Zn--Cl coordination environments remained stable across iterations.

\subsection{MLIP Training Protocol}
\label{sec:mlip_protocol}

\subsubsection{All-Atom Training}

We trained all-atom MLIP models using DeePMD-kit.\cite{DeepMD-2018,DeepMD-2023}
Each model takes the atomic coordinates and chemical species as input
and predicts the total energy $E$ of the system and the force
$\mathbf{F}_i$ acting on each atom $i$. The models were optimized
against DFT reference energies and forces using an energy--force
loss of the form
\begin{equation}
    L(\mathbf{w}) =
    \frac{p_\epsilon}{N}\left|E_{\rm DFT}-E_{\mathbf{w}}\right|^2
    +
    \frac{p_f}{3N}\sum_i
    \left|\mathbf{F}^{\rm DFT}_i-\mathbf{F}^{\mathbf{w}}_i\right|^2 .
\end{equation}
Here, $\mathbf{w}$ denotes the trainable model parameters,
$E_{\rm DFT}$ and $E_{\mathbf{w}}$ are the DFT and MLIP-predicted
total energies, and $\mathbf{F}^{\rm DFT}_i$ and
$\mathbf{F}^{\mathbf{w}}_i$ are the corresponding forces on atom $i$.
The number of atoms is denoted by $N$, and $p_\epsilon$ and $p_f$
control the relative weights of the energy and force contributions
during training.

The all-atom MLIPs were trained to reproduce the DFT potential-energy surface. 
The DeepMD representation decomposes the total potential energy
into a sum of atomic contributions,
$E_{\rm MLIP} = \sum_i E_i(\mathcal{D}_i)$,
where $\mathcal{D}_i$ is a smooth descriptor of the local environment
around atom $i$. The atomic forces are obtained analytically from
the gradient of the predicted total energy,
\begin{equation}
\mathbf{F}_i=-\nabla_{\mathbf{R}_i}E_{\rm MLIP}(\mathbf{R}),
\label{eq:mlip_force}
\end{equation}
where $\mathbf{R}$ denotes all atomic positions and $\mathbf{R}_i$
is the position of atom $i$.
It is worth stressing that although DeepMD uses local atomic descriptors, long-range Coulomb contributions are included in the DFT reference energies and forces, allowing their effects to be learned implicitly by the MLIP.

The DeepMD architecture contains two neural-network
components: an embedding network and a fitting network as shown later in Fig. \ref{fig:validation}. The embedding
network maps raw local structural information, including interatomic
distances and angular correlations, into symmetry-preserving
descriptors of each atomic environment. These descriptors are then
passed to the fitting network, which predicts the corresponding
atomic energy contributions. This construction preserves translational,
rotational, and permutational invariance while allowing the model
to learn many-body interactions from the DFT reference data.

For the all-atom MLIPs, we used the Deep Potential Smooth Edition
(DeepPot-SE) descriptor, \texttt{se\_a}, which incorporates radial
and angular local information, with a cutoff radius of 6.0~\AA{}
and a smoothing onset of 0.5~\AA{}. The embedding network contained
layers of 20, 40, and 80 neurons, with an axis embedding dimension
of 12, while the fitting network comprised four hidden layers of
240 neurons each. The final active-learning dataset contained
2886 DFT-labeled configurations. Each final DeepMD-kit v3.0 model
was trained for $2.0\times 10^6$ steps with a learning rate
decaying from $10^{-3}$ to $10^{-6}$, using a decay interval
of 5000 steps. The trained models were subsequently frozen and
compressed for production simulations.

\subsubsection{Coarse-Grained Training and Simulations for All-Atom Initialization}
\label{sec:cg_md}

To assess the convergence of Zn--Cl speciation with respect to
initial configurations in subsequent MLIP-MD simulations, we
developed an auxiliary ion-only coarse-grained (CG) potential
using DeePMD-kit.
The reference data were obtained from atomistic simulations
based on revPBE without D3 dispersion corrections, for a
3~m aqueous ZnCl$_2$ solution containing 26 Zn ions,
52 Cl ions, and 481 water molecules in a periodic cubic cell
of approximately 25.5~\AA{} per side.
We downselected 25,000 configurations using uniformly spaced
trajectory windows to reduce temporal correlations.
Only Zn and Cl coordinates were retained, while their reference
forces included interactions with all atoms, including water.
Solvent effects were thus incorporated approximately through
force matching.

The CG model employed the same DeepPot-SE descriptor and
neural-network architecture as the all-atom MLIPs, but with
a larger cutoff radius of 12~\AA{}.
Training was performed for $10^6$ steps using only ionic
forces, with energy and virial weights set to zero.
From the CG-MD simulations, we selected five configurations
(C2--C6) exhibiting distinct ionic structures and Zn--Cl
coordination environments. These configurations were solvated with water using
Packmol~\cite{Packmol-2009} and used to initialize
independent explicit-solvent DeepMD simulations lasting
12.5--25~ns to assess the sensitivity of the predicted
Zn--Cl speciation to the initial configurations. Further details and structural
comparisons with all-atom simulations are provided in the
Supporting Materials (SM).

\subsection{MLIP MD Simulations}
\label{MD_detail}

MLIP-driven MD simulations of aqueous \ce{ZnCl2} solutions from 1 to 30~m were performed using the revPBE-based MLIPs developed in this work and a previously reported SCAN-based MLIP.\cite{Cao-2025}
Production simulations at $3$--$30$ m were performed in the $NpT$ ensemble at 298 K and 1 bar using a timestep of 0.5 fs. The $NpT$ ensemble was chosen to allow the density to equilibrate, given the greater uncertainty in solution densities at these concentrations. Simulations at $1$--$2$ m were performed in the $NVT$ ensemble. Periodic boundary conditions were applied in all directions. Each system was simulated for approximately $20$--$50$ ns, depending on concentration, with $50$ ns trajectories used at higher concentrations. The first $10$--$25$ ns of each trajectory were discarded as equilibration, leaving $10$--$25$ ns of production data for the speciation and EXAFS analyses. These long trajectories are used to obtain reliable speciation statistics, with convergence demonstrated later. Representative simulation cells are summarized in Table~\ref{sim_data}.

\begin{table}[tbh]
\centering
\caption{Representative production simulation cells used for aqueous \ce{ZnCl2}. At 3~m, initial configurations were generated using either coarse-grained simulations (C2--C6) or Packmol\cite{Packmol-2009} to demonstrate convergence of the speciation statistics from different starting configurations.}
\begin{tabular}{ccccc}
\hline
$\#$ \ce{Zn^{2+}} & $\#$ \ce{H2O} & $\bar{l}$ [\AA]
& $c$ [mol\,L$^{-1}$] & $m$ [mol\,kg$^{-1}$] \\
\hline
$5$   & $267$ & $20$ & $1.0$  & $1.0$  \\
$10$  & $267$ & $20$ & $2.1$  & $2.1$  \\
$14$  & $259$ & $20$ & $2.9$  & $3.0$  \\
$26$  & $481$ & $25$ & $2.8$  & $3.0$  \\
$24$  & $216$ & $20$ & $5.0$  & $6.2$  \\
$62$  & $344$ & $25$ & $6.4$  & $10.0$ \\
$108$ & $200$ & $25$ & $10.6$ & $30.0$ \\
\hline
\end{tabular}
\label{sim_data}
\end{table}

\begin{figure}[tbh]
    \centering
    \includegraphics[width=\columnwidth]{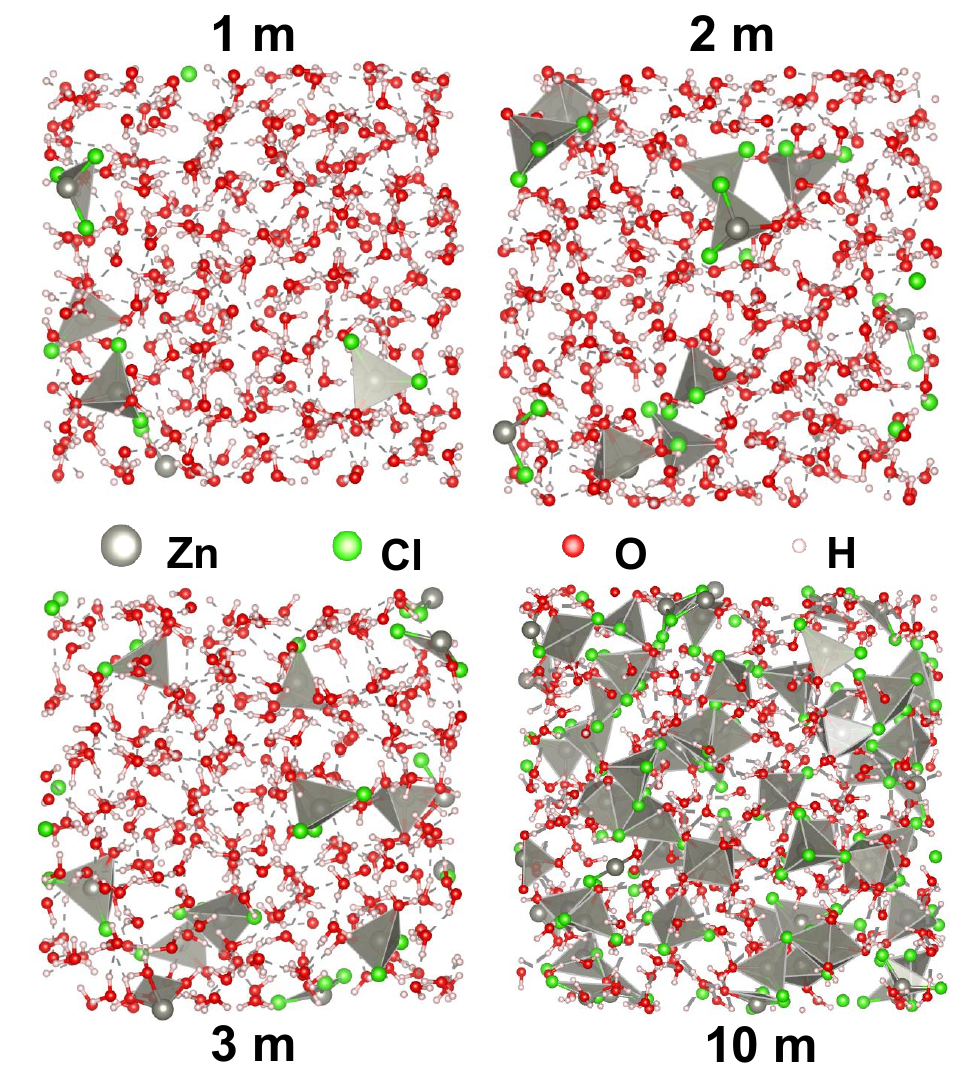}
    \caption{Representative configurations from production MLIP-MD simulations of aqueous \ce{ZnCl2} at the indicated concentrations. Increasing concentration promotes chloride coordination and the formation of connected Zn--Cl structures through shared chloride ligands. Zn-centered coordination polyhedra are shown in gray.}
    \label{fig:MD-configurations}
\end{figure}

Representative configurations from the production MLIP-MD trajectories are shown in Figure~\ref{fig:MD-configurations}. At low concentrations, Zn-centered coordination environments are predominantly isolated and hydrated by water. As concentration increases, chloride coordination becomes more prevalent, and neighboring Zn centers increasingly share chloride ligands, forming more connected Zn--Cl structures. These structural changes are examined quantitatively in the Results and Discussion.

Separately, Zn--Cl coordination-number-based potentials of mean force (PMFs) were calculated using umbrella-sampling MD simulations implemented through PLUMED.\cite{PLUMED}
The cubic simulation cell had a side length of $20$~\AA\ and contained one Zn ion, four Cl ions, and 264 water molecules. We computed the PMF along the continuous Zn--Cl coordination number, $n_{\rm Cl}$, defined as
\begin{equation}
n_{\rm Cl}
=
\sum_{j\in\{\mathrm{Cl}\}}
\frac{
1-\left[\left(r_{\mathrm{ZnCl},j}-d_0\right)/r_0\right]^n
}{
1-\left[\left(r_{\mathrm{ZnCl},j}-d_0\right)/r_0\right]^m
},
\label{eq:ncl}
\end{equation}
where $r_{\mathrm{ZnCl},j}$ is the instantaneous distance between the selected Zn ion and chloride ion $j$. The switching-function parameters were $r_0=3.0$~\AA, $d_0=0$, $n=10$, and $m=30$. A total of 51 umbrella windows with centers spanning $n_{\rm Cl}=0.0$--$4.0$ were employed. In window $i$, the coordination number was restrained using $U_i(n_{\rm Cl})=\frac{1}{2}\kappa(n_{\rm Cl}-n_i)^2$, with custom $\kappa$ at various windows. Check Table S2 of SM for detailed window-specific $\kappa$. Each window was simulated in the $NVT$ ensemble at $298.15$~K for $0.1$~ns. The PMF was reconstructed using the weighted histogram analysis method (WHAM) after discarding the initial $0.01$~ns of each trajectory for equilibration. The WHAM consistency checks are presented in the SM.

\subsection{Reference AIMD simulations}
\label{Sec:AIMD}

Ab initio molecular dynamics (AIMD) simulations of dilute aqueous \ce{ZnCl2} systems were performed using the Quickstep module of CP2K.\cite{CP2K-2020} 
For the coordination-number PMF calculations, the system contained one \ce{Zn^{2+}} ion, four \ce{Cl^-} ions, and 264 water molecules consistent with MLIP MD simulations: all systems were simulated in a cubic cell with a side length of 20 \AA{} under periodic boundary conditions.
Valence electrons were treated explicitly within density functional theory using the revised Perdew--Burke--Ernzerhof functional (revPBE)\cite{zhang1998comment} and the DZVP--MOLOPT--SR--GTH basis set,\cite{molopt} with a plane-wave density cutoff of 400 Ha. Core electrons were represented using Goedecker--Teter--Hutter pseudopotentials,\cite{GTH1} and long-range dispersion interactions were included through the Grimme DFT-D3 correction.\cite{grimme2010consistent}

The simulations were performed in the canonical ensemble at 300 K using a Nos$\rm{\acute{e}}$-Hoover thermostat. The umbrella windows were separated by 0.1 in coordination number. A time step of 0.5 fs was used, and each window was simulated for 20 ps. The corresponding restraint strengths were adjusted according to the position of each window relative to the local free-energy minima. The first 2.5 ps were discarded as equilibration, and the remaining 17.5 ps were used for analysis.

\subsection{Classical MD simulations}
\label{Sec:CMD}

To provide a comparison with the MLIP simulations, classical MD simulations of aqueous \ce{ZnCl2} solutions were performed using the LAMMPS software package\cite{lammps-2022}. Water was represented by the SPC/E model,\cite{Straatsma-1987} while the \ce{Zn^{2+}}--\ce{Cl^-} and ion--water interactions were described using the pair-specific Lennard--Jones parameters of a carefully developed and experimentally benchmarked classical
force field for aqueous \ce{ZnCl2} by Stack and co-workers.
\cite{Rampal-2021,Dhakal2023ZnHalides}
This force field was specifically refined against neutron
diffraction measurements. The SPC/E O--H bonds and H--O--H angles were constrained using the SHAKE algorithm with a relative tolerance of $10^{-4}$. Lennard--Jones and real-space electrostatic interactions were truncated at 9~\AA, while long-range electrostatic interactions were evaluated using the particle--particle particle--mesh method with a target relative force accuracy of $10^{-4}$. Simulations were performed under periodic boundary conditions in the isothermal--isobaric ensemble at 298~K and 1~atm using the Nos$\rm{\acute{e}}$--Hoover thermostat and barostat implemented, with relaxation times of 50 and 500~fs, respectively. A cell-binning Verlet neighbor list with a 2.0~\AA\ skin distance was updated every step when required, and the equations of motion were integrated using a timestep of 1~fs. The input shown corresponds to a 1~ns production segment.

\subsection{Speciation Analysis}
\label{sec:speciation}

The populations of Zn--chloride complexes were determined from the instantaneous first coordination shell of each Zn ion. For every trajectory frame, a molecular graph was constructed by assigning Zn--Cl and Zn--O contacts when the corresponding interatomic distance was less than $3.0$ \AA . An O--H cutoff of $1.2$ \AA\ was used to preserve the identity of individual water molecules. Each Zn ion was then classified according to the numbers of directly coordinated chloride ions, $n$, and water molecules, $m$, and the resulting species were denoted as \ce{[ZnCl_n(H2O)_m]^{2-n}}. Species populations were obtained by counting the occurrence of each coordination-shell composition over all Zn ions and trajectory frames and normalizing by the total number of Zn-ion--frame observations. This procedure provides the relative population of each Zn coordination environment at a given electrolyte concentration. In what follows, for 3~m we obtain the uncertainties (standard deviations) in speciation from independent MD simulations while for other concentrations, the block averages from one long trajectory during the production are used.

\subsection{EXAFS Experimental Details}
\label{sec:expt_protocol}

Aqueous samples of ZnCl$_2$ (and LiCl in some cases) were prepared in air by dissolving weighed amounts of the salts in HPLC‑grade water. A magnetic stirrer was used for the high‑molarity samples. Visual inspection indicated that the solutions were clear, with no evidence of undissolved solute. All reagents were purchased from standard vendors (Alfa Aesar, Sigma‑Aldrich, Fisher Scientific) at 99.9\% purity or higher.

X-ray absorption spectroscopy was performed at beamline 20-BM of the Advanced Photon
Source. Aqueous solutions of ZnCl$_2$ in varying concentrations were placed in PEEK cells with polyimide windows and sealed with epoxy. XAS measurements were acquired in transmission mode by placing gas ionization chambers before (50\% He, 50\% N$_2$) and after (100\% N$_2$) each sample. The incident beam was reduced to a spot size of 1 mm horizontal by 0.5 mm vertical by a toroidal mirror. Harmonic rejection was performed using a Rh-coated flat mirror and detuning the incident beam by 15\%. Spectra were acquired at the Zn K-edge and all spectra were energy aligned to that of the first derivate peak of a Zn reference foil (9660.76 eV).\cite{Kraft-1996} Spectral background removal and normalization of the XAS spectra were performed within the Athena software program.\cite{Ravel-2005} Specifically, the EXAFS data was extracted through the use of a spline function and a cutoff distance (Rbkg) of 1.0 \AA .

\subsection{Theoretical EXAFS Calculations}
\label{sec:theoretical_exafs}

Theoretical Zn K-edge EXAFS spectra were calculated using FEFF10\cite{kas2021advanced} from Zn-centered clusters extracted from 5,000 representative snapshots of DPMD trajectories generated with MLIPs trained at their respective DFT reference levels. The ensemble-averaged EXAFS signal was calculated as
\begin{equation}
\overline{\chi}(k)
=
\frac{1}{N_{\mathrm{env}}}
\sum_{i=1}^{N_{\mathrm{env}}}
\chi_i(k),
\label{eq:exafs-config-average}
\end{equation}
where $N_{\mathrm{env}}$ is the number of sampled Zn-centered environments and $\chi_i(k)$ is the EXAFS function for the $i$th environment. This averaging accounts for structural fluctuations, including Zn--Cl--Zn bridging configurations, without introducing additional fitted Debye--Waller factors.
Identical FEFF10 parameters,\cite{kas2021advanced} energy shifts, amplitude-reduction factors, and Fourier-transform settings were used for all three ensembles, ensuring that spectral differences reflect the underlying sampled structures rather than differences in FEFF10 settings. Further FEFF10 computational details and calculation parameters are provided in the Supplemental Material (SM).

\section{Results \& Discussion}
\label{sec:results_discussions}

In this section, we present the results of our molecular dynamics simulations and Zn K-edge EXAFS experiments on aqueous \ce{ZnCl2} solutions. We first validate the custom-trained MLIPs against DFT calculations and examine the convergence of the predicted speciation. We then investigate how local Zn coordination changes with concentration and compare the calculated EXAFS spectra with experiment to assess the effects of the exchange--correlation functional and dispersion treatment.
Next, we compare the custom-trained MLIPs with a universal MLIP and classical MD in terms of structural predictions and computational efficiency. Finally, we summarize the concentration-dependent coordination trends and discuss their implications for Zn--Cl speciation.

\begin{figure}[tbh]
    \centering

    \subfloat[MLIP model: DeepMD]{%
        \includegraphics[width=0.78\columnwidth]{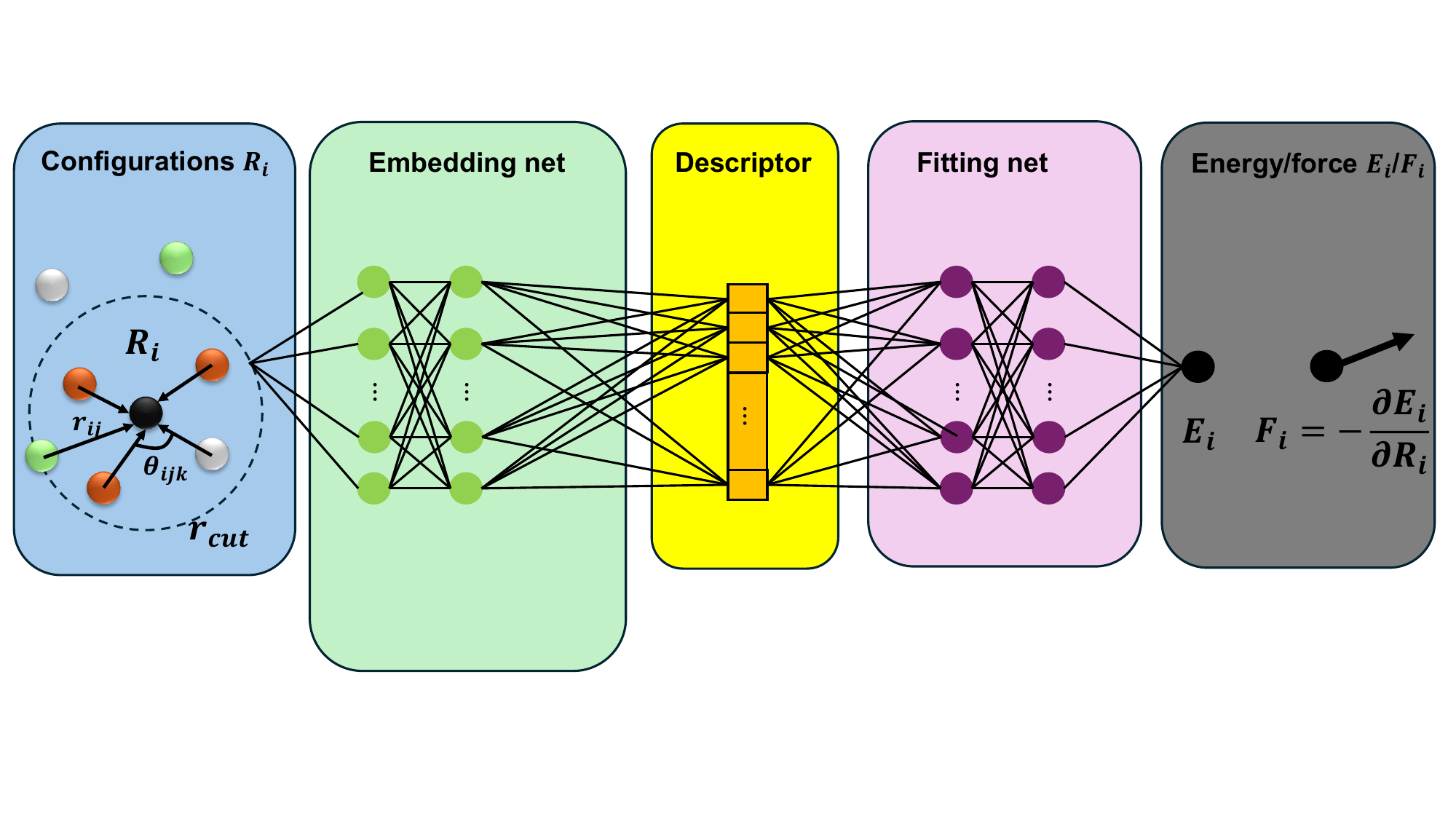}
        \label{fig:panel_a}
    }

    \vspace{-1mm}

    \subfloat[revPBE with D3 for all atoms]{%
        \includegraphics[width=0.78\columnwidth]{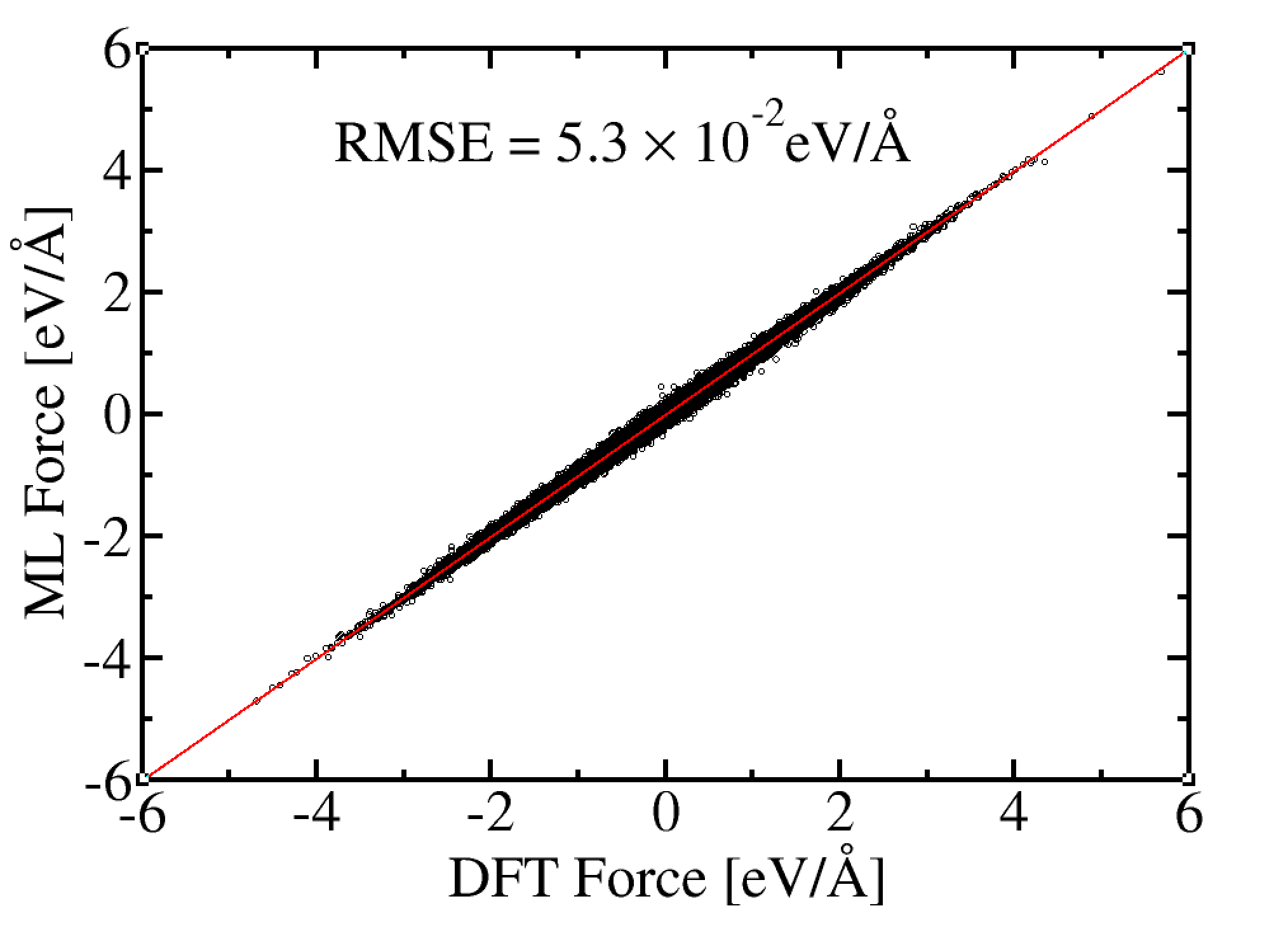}
        \label{fig:panel_b}
    }

    \vspace{-1mm}

    \subfloat[revPBE without D3 for \ce{Zn^{2+}} ions]{%
        \includegraphics[width=0.78\columnwidth]{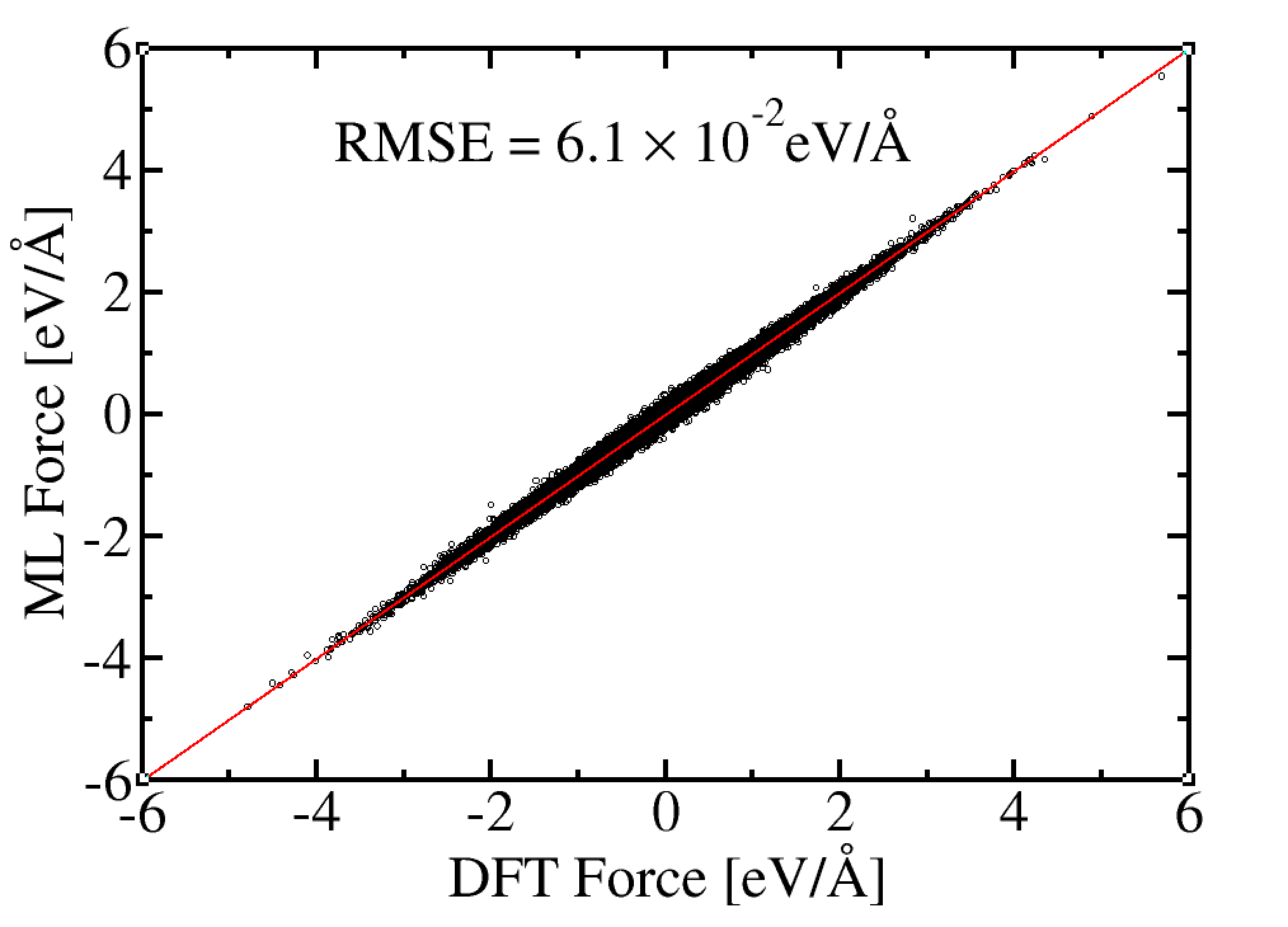}
        \label{fig:panel_c}
    }

    \caption{%
        (a) Schematic of the DeepMD model and representative MD snapshots
        of aqueous ZnCl$_2$ at different concentrations.
        MLIP forces compared with DFT calculations using
        (b) revPBE with D3 and (c) revPBE without D3 on Zn$^{2+}$ for randomly selected
        configurations. Force root-mean-square errors (RMSEs) are shown.
    }
    \label{fig:validation}
\end{figure}

\subsection{Validation of the Machine-Learning 
Interatomic Potentials}
\label{sec:mlip_validation}

An overview of the DeepMD architecture and representative aqueous \ce{ZnCl2} configurations is presented in Fig.~\ref{fig:validation}(a). The training protocol is described in Sec.~\ref{sec:mlip_protocol}. The MLIPs were validated by comparing their predictions with DFT reference energies and forces for independent configurations generated after training, as well as structural and coordination free-energy observables obtained from DFT simulations.

Figures~\ref{fig:validation}(b) and \ref{fig:validation}(c) compare the predicted and DFT reference forces for the two revPBE-based MLIPs, with each point representing one Cartesian force component. The predictions closely follow the ideal $y=x$ relationship, with root-mean-square errors (RMSEs) of $5.3\times10^{-2}$ and $6.1\times10^{-2}$~eV\,\AA$^{-1}$ for revPBE-D3 and revPBE-wo-D3-on-Zn, respectively. These comparable errors indicate that both MLIPs accurately reproduce their respective DFT reference forces. We next examine their ability to reproduce the DFT coordination free-energy landscape.

\begin{figure}[tbh]
    \centering
    \includegraphics[width=\columnwidth]{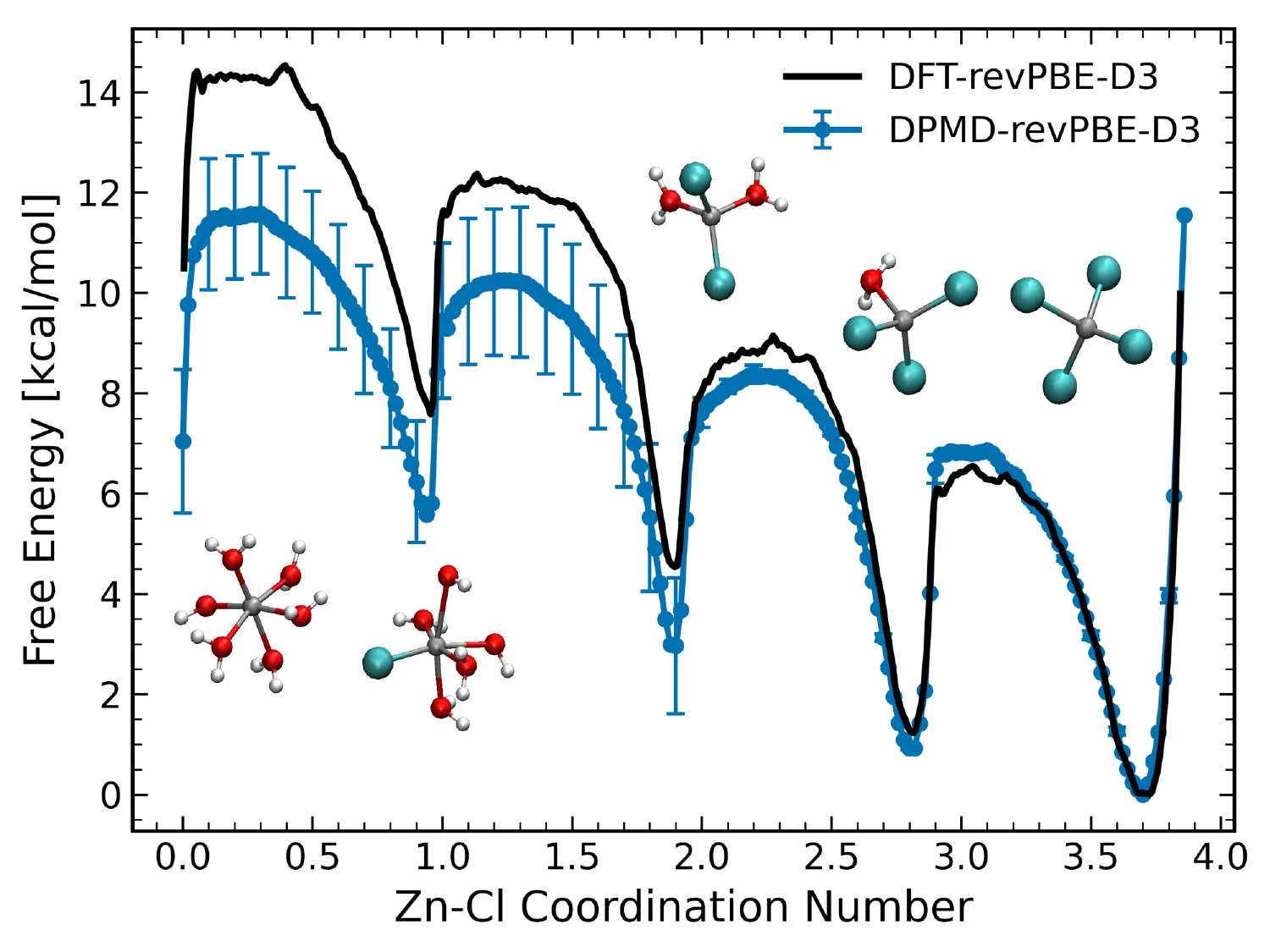}
    \caption{Validation of the Deep MD potential (DPMD) model against revPBE-D3 density functional theory (DFT) using the Zn--Cl coordination-number potential of mean force}
    \label{fig:validation-pmf}
\end{figure}

Figure~\ref{fig:validation-pmf} compares the Zn--Cl coordination-number potentials of mean force obtained from DPMD and direct revPBE-D3 DFT simulations. The revPBE-D3 results identify the four-chloride-coordinated Zn environment as the lowest free-energy state, with a minimum near a Zn--Cl coordination number of 3.7. The minima corresponding to three, two, and one chloride neighbors are approximately 1, 4.5, and 7.5~kcal/mol higher in free energy, respectively. The apparent free-energy barriers for increasing chloride coordination are approximately 4.5--5.5~kcal/mol. Overall, DPMD reproduces the positions and relative ordering of the principal free-energy minima, as well as the main features of the barriers separating them. Agreement is particularly good at higher coordination numbers, while DPMD predicts somewhat lower free energies for the one- and two-chloride coordination environments. These results demonstrate that DPMD captures the principal features of the revPBE-D3 coordination free-energy landscape. The remaining quantitative differences may partly arise from the more limited sampling accessible to direct DFT simulations, particularly for less frequently sampled coordination environments.

\subsection{Speciation and Sampling Convergence}
\label{sec:speciation_convergence}

\begin{figure*}[t]
    \centering
    \includegraphics[width=0.8\textwidth]{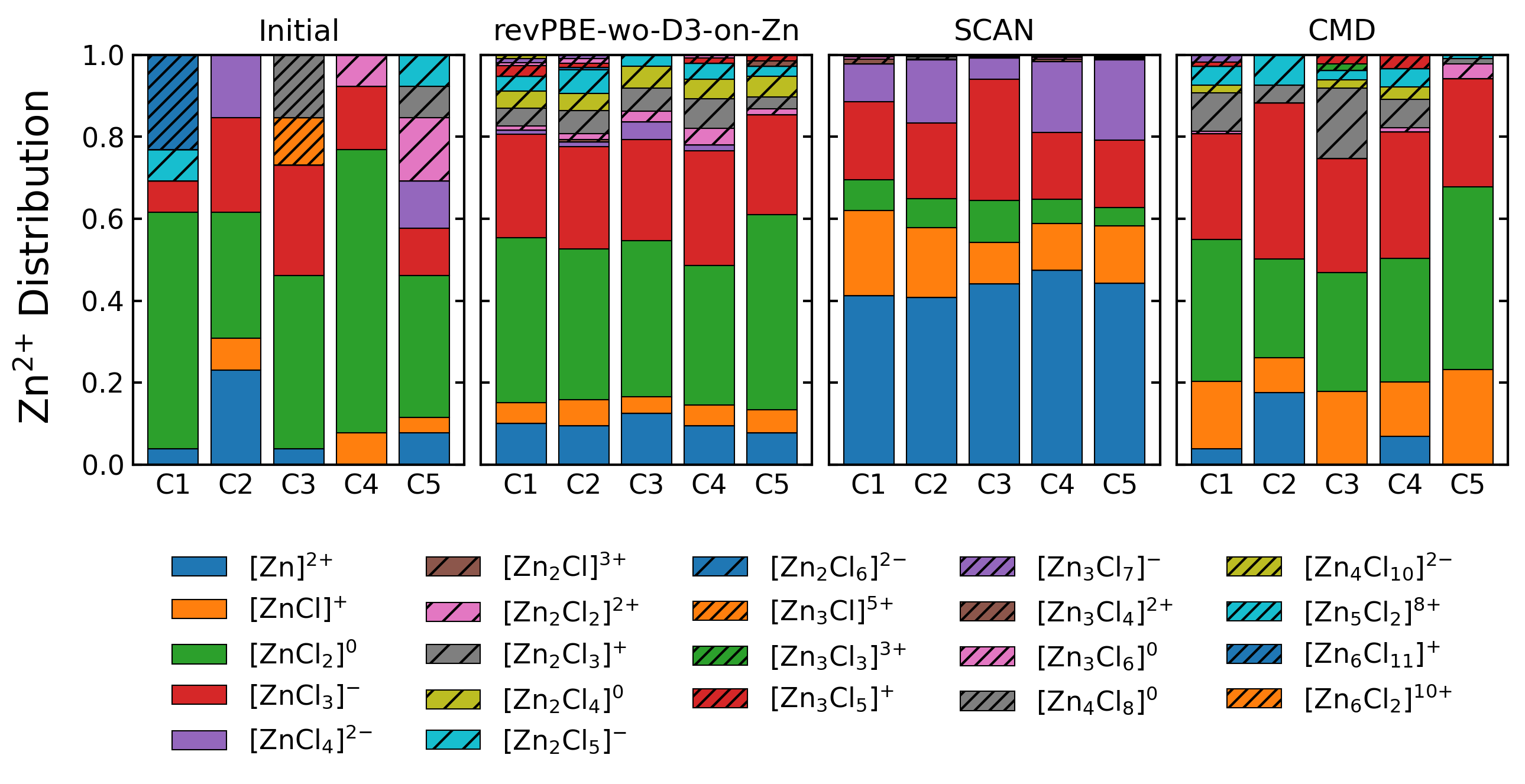}
    \caption{
    Speciation convergence analysis for aqueous \ce{ZnCl2}
    at 3~m using DPMD-revPBE without
    D3 on Zn, DPMD-SCAN, and classical MD (CMD).
    The y-axis denotes the average population of $\mathrm{Zn}^{2+}$ in each species, where each bar segment represents a unique $[\mathrm{Zn}_{n}\mathrm{Cl}_{m}]^{(2n-m)+}$ cluster in the legend.
    For each of the models, five simulations were initiated from the same set of five coarse-grained (C1--C5) configurations.
    DPMD-revPBE and DPMD-SCAN species distributions were averaged over 5~ns, whereas the CMD distributions were averaged over 50~ns.
    Both the DPMD-revPBE and DPMD-SCAN models exhibit relatively uniform distributions for their dominant species, suggesting that they have reached their respective equilibrium distributions.
    However, despite the $10\times$ larger window for averaging, the CMD simulations still display a larger variance in their species distributions.
    Representative DPMD EXAFS results for C3 are provided in SM.
    }
    \label{fig:speciation}
\end{figure*}

Previous work by Cao {\it et al.} employed three independent 100~ns MD trajectories at a wide range of concentrations to calculate properties of interest.\cite{Cao-2025} Here, to assess whether the predicted speciation at a given concentration depends on the initial Zn--Cl organization, we performed multiple independent DeepMD and CMD simulations at 3~m starting from substantially different configurations. Five initial configurations (C1--C5) were obtained from structurally distinct coarse-grained simulations and adjusted to the target composition by removing water molecules. An additional configuration was constructed independently using Packmol (PM).\cite{Packmol-2009}
These configurations were used for the
revPBE-wo-D3-on-Zn and SCAN simulations. As shown in
Figure~\ref{fig:speciation}, the initial configurations
exhibit markedly different species distributions,
including different populations of mononuclear Zn
complexes and multinuclear Zn--Cl aggregates.
These configurations therefore provide a stringent
test of whether the simulations retain memory of
their initial Zn--Cl organization.

The DeepMD trajectories were propagated for 25--50~ns each, with the initial 10--25~ns of each trajectory discarded as equilibration, and the species populations shown in Figure~\ref{fig:speciation} were obtained by averaging over all Zn ions and trajectory frames from 12.5 to 25~ns. Despite the substantial differences among the initial distributions, the simulations yield closely similar production-averaged populations of the dominant mononuclear species. In particular,
the relative populations of \ce{Zn^{2+}}, \ce{ZnCl+},
\ce{ZnCl2}, and \ce{ZnCl3-} become largely
independent of the initial configuration.
In contrast, the CMD simulations retain substantially larger differences in speciation among the initial configurations, even after 50~ns of equilibration followed by 50~ns of production sampling. This slower convergence is consistent with kinetic trapping associated with strong fixed point-charge electrostatic interactions, which are not explicitly imposed in our second-generation DeepMD potentials.

Some variation remains among the less abundant
multinuclear Zn--Cl aggregates, whose populations
are more sensitive to finite sampling. Nevertheless,
the substantial loss of memory of the initial
configurations indicates that the dominant species
populations at 3~m are adequately equilibrated
for the DeepMD speciation analysis presented here.
This distinction is important because convergence
of the dominant local coordination environments
does not necessarily imply complete sampling of
all possible multinuclear structures.

\begin{table}[tbh]
    \centering
    \caption{Fractions of Zn ions classified as
    mononuclear Zn--Cl complexes or multinuclear
    aggregates at different \ce{ZnCl2} concentrations,
    obtained using SCAN and revPBE-wo-D3-on-Zn.
    Here, $n_{\mathrm{Zn}}$ denotes the number of Zn
    ions in an aggregate. Subscripts indicate standard
    deviations in the last reported digits where
    available.}
    \label{tab:zncl-speciation-polymer}
    \small
    \setlength{\tabcolsep}{4pt}
    \begin{tabular}{lcccc}
        \hline
        Species/category & 1~m & 3~m & 10~m & 30~m \\
        \hline
        \multicolumn{5}{l}{\textbf{SCAN}} \\
        \ce{Zn^{2+}}    & 0.842 & $0.435_{24}$ & 0.305 & 0.064 \\
        \ce{ZnCl^{+}}   & 0.098 & $0.147_{39}$ & 0.059 & 0.061 \\
        \ce{ZnCl2}     & 0.009 & $0.071_{19}$ & 0.058 & 0.053 \\
        \ce{ZnCl3^{-}}  & 0.051 & $0.200_{49}$ & 0.241 & 0.061 \\
        \ce{ZnCl4^{2-}} & 0.000 & $0.133_{54}$ & 0.205 & 0.023 \\
        $n_{\mathrm{Zn}}=2$    & 0.000 & $0.014_{05}$ & 0.126 & 0.142 \\
        $n_{\mathrm{Zn}}\geq 3$ & 0.000 & 0.000 & 0.007 & 0.596 \\
        \hline
        \multicolumn{5}{l}{\textbf{revPBE-wo-D3-on-Zn}} \\
        \ce{Zn^{2+}}    & 0.250 & $0.097_{14}$ & 0.003 & 0.001 \\
        \ce{ZnCl^{+}}   & 0.016 & $0.049_{09}$ & 0.003 & 0.002 \\
        \ce{ZnCl2}     & 0.246 & $0.400_{48}$ & 0.021 & 0.017 \\
        \ce{ZnCl3^{-}}  & 0.368 & $0.254_{11}$ & 0.013 & 0.009 \\
        \ce{ZnCl4^{2-}} & 0.073 & $0.013_{14}$ & 0.000 & 0.000 \\
        $n_{\mathrm{Zn}}=2$    & 0.047 & $0.156_{28}$ & 0.185 & 0.034 \\
        $n_{\mathrm{Zn}}\geq 3$ & 0.000 & $0.031_{15}$ & 0.775 & 0.937 \\
        \hline
    \end{tabular}
\end{table}

\begin{table}[tbh]
    \centering
    \caption{Fractions of Zn ions having different numbers of first-shell Cl neighbors ($n_{\mathrm{Cl}}$) at different \ce{ZnCl2} concentrations, obtained using SCAN and revPBE-wo-D3-on-Zn. Here, $n_{\mathrm{Cl}}$ denotes the number of Cl ions coordinated to a Zn ion, regardless of whether it belongs to a larger aggregate. Subscripts indicate standard deviations in the last reported digits where available.}
    \label{tab:zncl-speciation}
    \small
    \setlength{\tabcolsep}{4pt}
    \begin{tabular}{lcccc}
        \hline
        $n_{\mathrm{Cl}}$ & 1~m & 3~m & 10~m & 30~m \\
        \hline
        \multicolumn{5}{l}{\textbf{SCAN}} \\
        0 & 0.842 & $0.436_{24}$ & 0.307 & 0.064 \\
        1 & 0.098 & $0.155_{41}$ & 0.100 & 0.093 \\
        2 & 0.009 & $0.073_{19}$ & 0.079 & 0.129 \\
        3 & 0.051 & $0.201_{49}$ & 0.278 & 0.279 \\
        4 & 0.000 & $0.135_{54}$ & 0.236 & 0.435 \\
        \hline
        \multicolumn{5}{l}{\textbf{revPBE-wo-D3-on-Zn}} \\
        0 & 0.229 & $0.098_{14}$ & 0.030 & 0.003 \\
        1 & 0.044 & $0.080_{09}$ & 0.080 & 0.014 \\
        2 & 0.261 & $0.463_{41}$ & 0.390 & 0.131 \\
        3 & 0.401 & $0.331_{26}$ & 0.417 & 0.397 \\
        4 & 0.064 & $0.029_{15}$ & 0.082 & 0.455 \\
        \hline
    \end{tabular}
\end{table}

We further quantify the concentration dependence of
Zn--Cl speciation using two complementary classifications.
Table~\ref{tab:zncl-speciation-polymer} distinguishes
mononuclear complexes from multinuclear aggregates.
For revPBE-wo-D3-on-Zn, the fraction of Zn ions
belonging to aggregates containing three or more Zn
ions increases to approximately 0.78 at 10~m and
0.94 at 30~m. In contrast,
Table~\ref{tab:zncl-speciation} classifies each Zn ion
according to the number of Cl neighbors in its first
coordination shell, irrespective of its connectivity
to other Zn ions. Consequently, the local Zn--Cl
coordination fractions should not be interpreted
as populations of distinct mononuclear complexes,
particularly at high concentrations where
chloride-bridged structures become prevalent.

Therefore, our independent-initial-configuration analysis
provides a complementary assessment of speciation
convergence, rather than relying on trajectory
length alone. Having established the robustness
of the dominant populations at 3~m, we next
examine the concentration dependence of local
structure and the coordination free-energy
landscapes predicted by different DFT descriptions.

\subsection{Local Structure and Coordination}

\begin{figure}[tbh]
    \centering
    \includegraphics[width=\columnwidth]{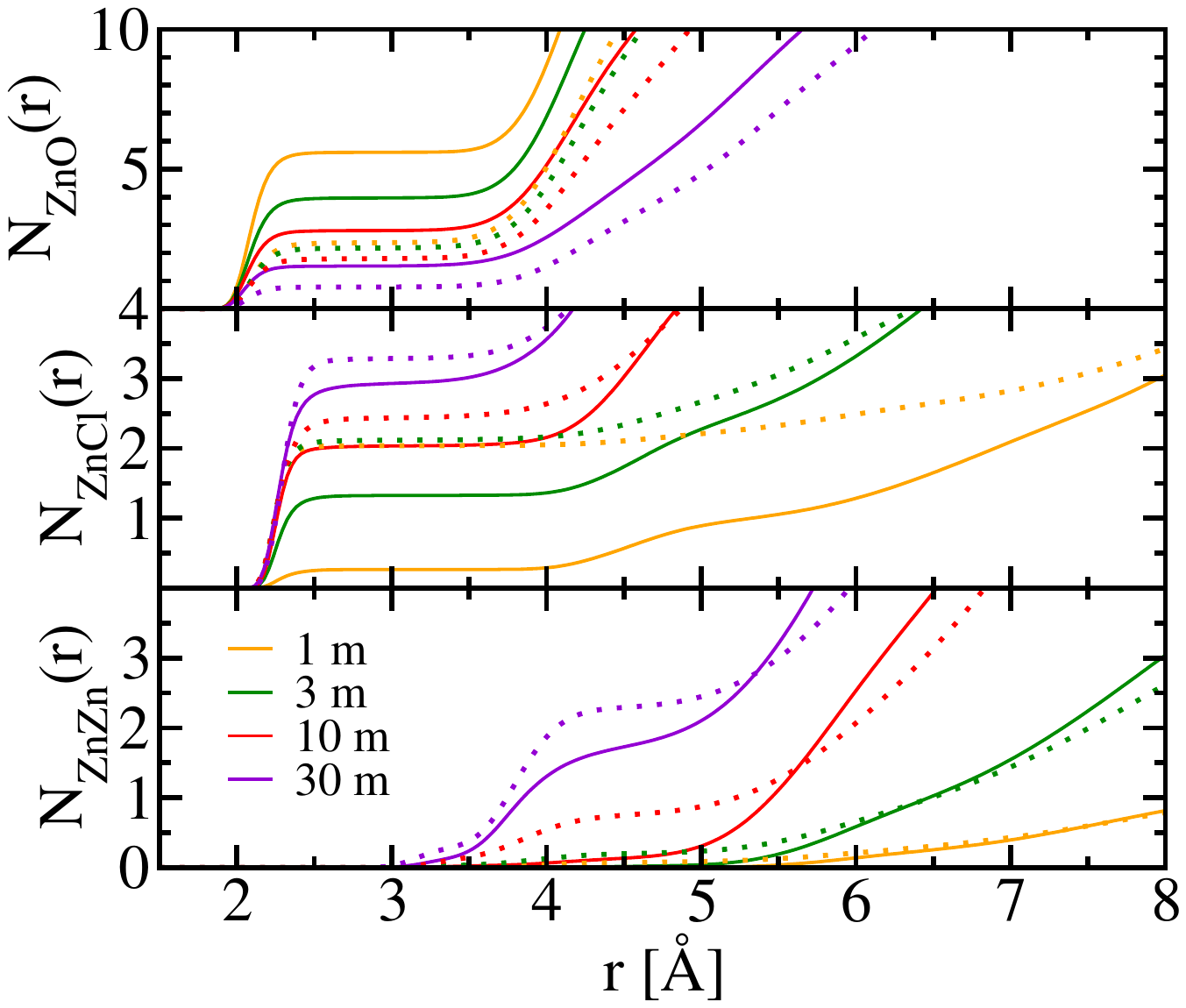}
    \caption{Running coordination numbers for Zn--O,
    $N_{\mathrm{ZnO}}(r)$ (top), Zn--Cl,
    $N_{\mathrm{ZnCl}}(r)$ (middle), and Zn--Zn,
    $N_{\mathrm{ZnZn}}(r)$ (bottom), in aqueous \ce{ZnCl2}
    solutions at 1, 3, 10, and 30~m. Solid and dotted lines
    correspond to MD simulations using MLIPs trained at
    the SCAN and revPBE levels, respectively. For the
    revPBE reference calculations, D3 dispersion
    contributions involving \ce{Zn^{2+}} were excluded
    (revPBE-wo-D3-on-Zn). Colors indicate molality.}
    \label{fig:NUM-conc}
\end{figure}

Figure~\ref{fig:NUM-conc} shows the running coordination numbers obtained using MLIPs trained with SCAN (solid lines) and revPBE (dashed lines), with the D3 dispersion corrections involving Zn ions excluded in the latter (revPBE-wo-D3-on-Zn).  They are obtained by integrating the corresponding radial distribution functions (RDFs), which are provided in Figure S6 of SM. 
The RDFs show that the first-shell Zn--O and Zn--Cl peak positions remain near 2.1 and 2.3~\AA, respectively,
with only modest variations across concentrations and between the two MLIPs. In contrast, the peak intensities and correlations beyond the first shell exhibit more pronounced differences, particularly the Zn--Zn correlations at higher concentrations.
Each running coordination number gives the mean
number of neighbors within a distance $r$ of a Zn ion.

Increasing concentration decreases the first-shell Zn--O coordination and increases the Zn--Cl
coordination for both aforementioned MLIPs, consistent with the progressive replacement
of water by chloride in the local Zn environment.
For SCAN, the first-shell Zn--O coordination decreases
from approximately $\sim$5.5 at 1~m to $\sim$1.5 at 30~m, while
the Zn--Cl coordination increases from approximately
0.3 to $\sim$3. In comparison, revPBE-wo-D3-on-Zn predicts
lower Zn--O coordination, decreasing from approximately
$\sim$2.5 to 0.8, and higher Zn--Cl coordination, increasing
from approximately 2.0 to 3.3 over the same concentration
range. The differences between the two MLIPs are particularly pronounced at lower concentrations. Therefore, at the same
concentration, SCAN yields higher water coordination
and lower chloride coordination than revPBE-wo-D3-on-Zn.
The differences are particularly pronounced at 1 and
3~m, indicating that the balance between water and
chloride coordination remains sensitive to the
underlying DFT reference description.

The bottom panel of Figure \ref{fig:NUM-conc} shows increasing short-range Zn--Zn coordination at high concentrations, most clearly at
30~m, where both models exhibit a pronounced rise
near $r=3.5$--$4.5$~\AA. The revPBE-wo-D3-on-Zn model
predicts more nearby Zn ions than SCAN, with a
particularly clear difference at 10~m. These correlations
are consistent with increased formation of chloride-bridged
multinuclear structures. 

\begin{figure}[tbh]
    \centering
    \includegraphics[width=\columnwidth]{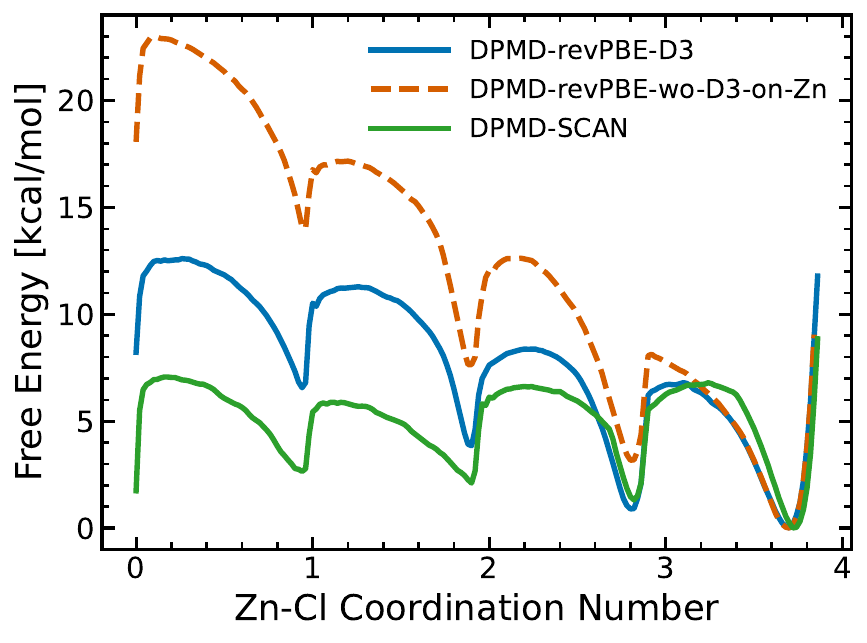}
    \caption{Zn--Cl coordination-number (CN) potentials
    of mean force calculated using MLIPs trained with
    different exchange--correlation functionals and
    dispersion treatments: revPBE-D3, revPBE-wo-D3-on-Zn,
    and SCAN.}
    \label{fig:functional-cl-cn}
\end{figure}

The differences in local Zn coordination between SCAN and revPBE-wo-D3-on-Zn suggest that the underlying DFT treatment influences the relative stability of Zn--Cl coordination states.
To examine this further, Figure~\ref{fig:functional-cl-cn}
compares the Zn--Cl coordination-number (CN) potentials
of mean force (PMFs) obtained using three MLIPs trained at the SCAN, revPBE-D3, and revPBE-wo-D3-on-Zn levels noting that comparing the two revPBE-based models allows us to
assess the effect of dispersion corrections involving Zn. As can be seen,
all PMFs predict the fourfold chloride-coordinated Zn state
as the most stable, corresponding to a ZnCl$_4$-like coordination environment. The relative free energies of the other coordination states, however, depend
on the underlying DFT treatment.

Taking the minimum near
coordination number 4 as the reference, the free-energy
minimum near coordination number 1 is approximately
3, 7, and 14--15~kcal/mol higher for SCAN,
revPBE-D3, and revPBE-wo-D3-on-Zn, respectively.
At coordination number 2, the corresponding free
energies are approximately 2, 4, and 8~kcal/mol.
These differences indicate that lower chloride
coordination states are energetically more accessible
with SCAN, whereas revPBE-wo-D3-on-Zn strongly
favors higher chloride coordination.

The free-energy barriers between adjacent coordination
states also depend on the functional and dispersion
treatment. For SCAN, the barriers for increasing Zn--Cl
coordination from 1 to 2, 2 to 3, and 3 to 4 are
approximately 3, 4.5, and 5.5~kcal/mol, respectively.
The corresponding barriers are approximately
5, 4.5, and 6~kcal/mol for revPBE-D3 and
3, 5, and 5~kcal/mol for revPBE-wo-D3-on-Zn.
Thus, while the barriers vary moderately among
the three methods, the differences in the relative
free energies of the coordination states are more
pronounced. These barriers are projected along
the coordination-number coordinate and do not
necessarily represent the complete kinetics of
ligand exchange. The greater energetic preference
for highly chlorinated environments predicted
by revPBE-wo-D3-on-Zn is consistent with the
higher Zn--Cl and lower Zn--O coordination numbers
observed in Figure~\ref{fig:NUM-conc}.

The CN PMFs also suggest a possible route toward
predicting concentration-dependent coordination
populations without requiring extensive sampling
at every concentration. In future work, we aim to combine these free-energy landscapes with concentration-dependent thermodynamic driving forces and ion-association effects to predict local Zn--Cl coordination populations and use the resulting populations and structural information to calculate EXAFS spectra across a broad range of concentrations.
Such an approach could reduce the need for
separate long MD simulations or enhanced-sampling
calculations at each concentration, although
its applicability to highly concentrated solutions
containing chloride-bridged multinuclear structures
remains to be established.

\subsection{EXAFS Across a Wide Range of Concentrations}

\begin{figure}[!tbh]
    \centering
    \includegraphics[width=0.95\columnwidth]{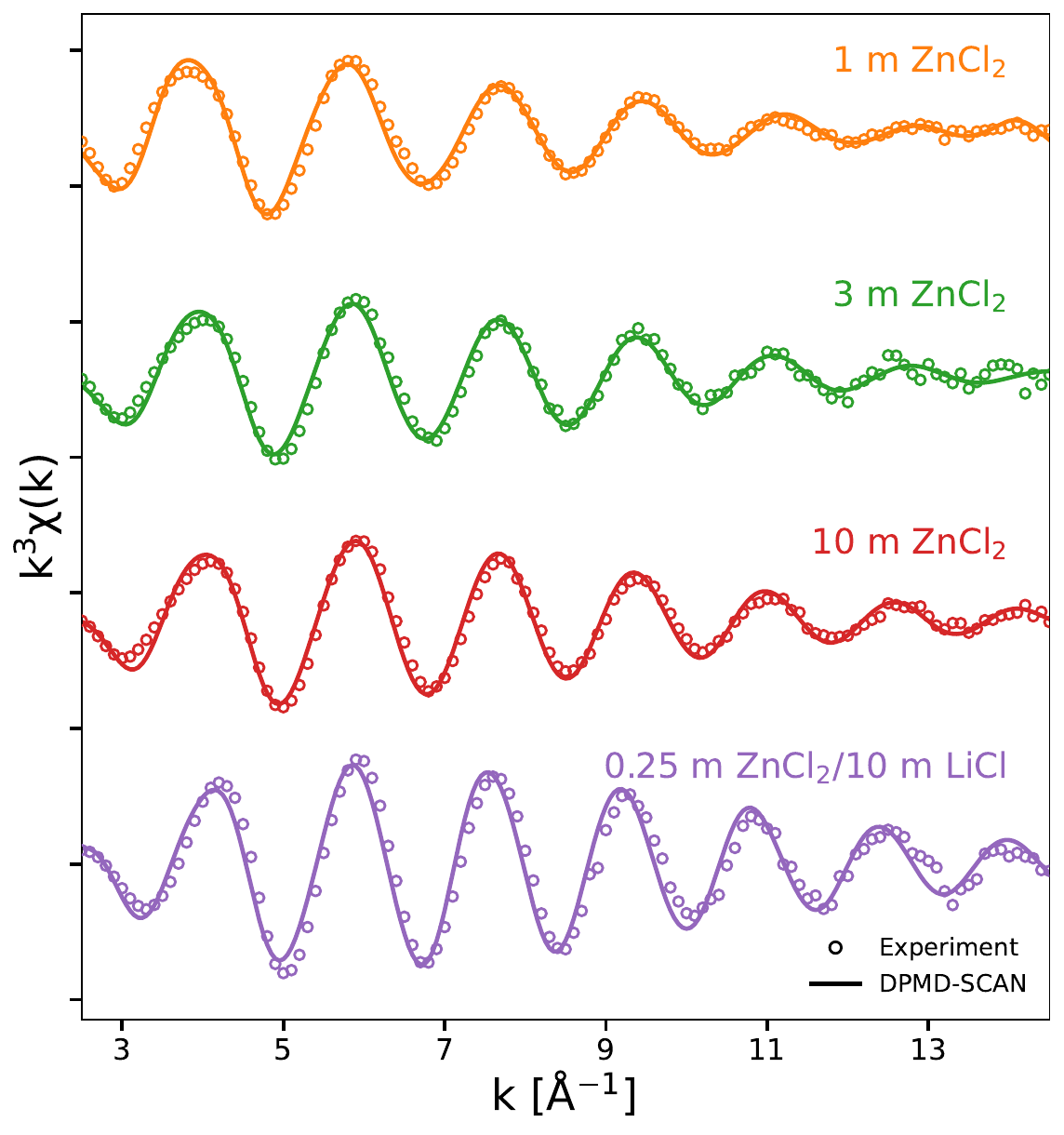}

    \vspace{0.3cm}

    \includegraphics[width=0.95\columnwidth]{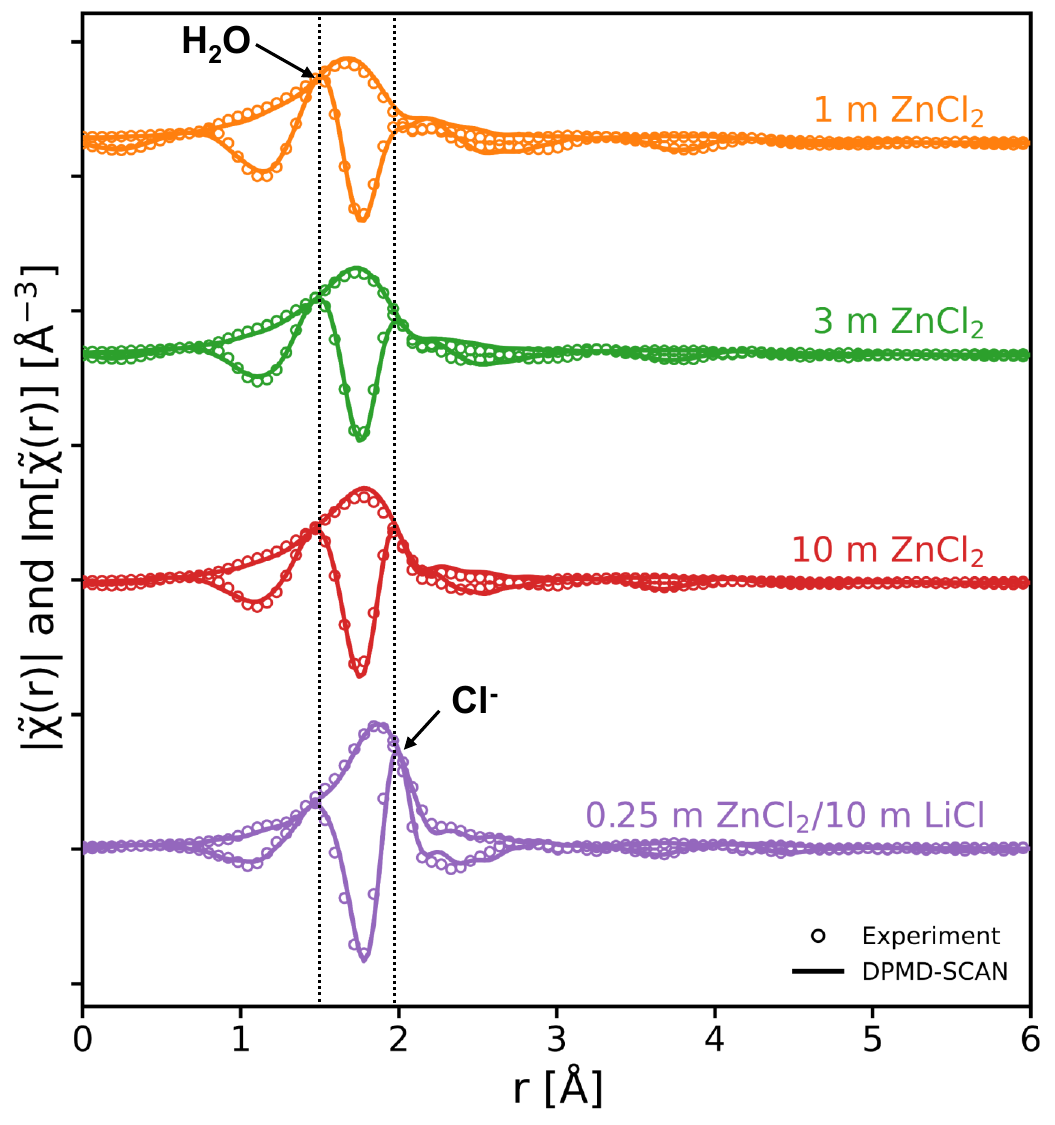}

    \caption{Comparison of experimental (open circles) and
    calculated (solid lines) Zn K-edge EXAFS spectra
    obtained using SCAN-based DPMD simulations for
    aqueous \ce{ZnCl2} solutions at different concentrations,
    including a mixed \ce{ZnCl2}/LiCl electrolyte.
    The upper and lower panels show the $k^3\chi(k)$ spectra
    and the magnitude and imaginary component of their
    Fourier transforms, respectively.
    Features near apparent distances of 1.5 and 2.0~\AA{}
    are associated primarily with Zn--O and Zn--Cl
    scattering, respectively. 
    Spectra are vertically offset for clarity.}
    \label{fig:exafs-concentration}
\end{figure}

Figure~\ref{fig:exafs-concentration} compares the
experimental Zn K-edge EXAFS spectra with those
calculated from SCAN-based DPMD simulations for
aqueous \ce{ZnCl2} solutions at different concentrations.
The calculated $k^3\chi(k)$ spectra reproduce the
principal experimental oscillations at 1, 3, and 10~m,
including changes in their phases and amplitudes
with increasing concentration. Although some differences
remain, particularly in the oscillation amplitudes at
larger $k$, the overall agreement indicates that the
simulations capture important features of the local
Zn coordination environments and their structural disorder.

The corresponding Fourier transforms provide further
insight into the concentration-dependent changes in Zn
coordination. The features near apparent distances of
1.5 and 2.0~\AA{} are associated primarily with Zn--O
and Zn--Cl scattering contributions, respectively.
As the \ce{ZnCl2} concentration increases, the dominant
peak in the Fourier-transform magnitude gradually shifts
toward larger apparent distances, consistent with an
increasing relative contribution from Zn--Cl coordination.
This trend is reproduced by the SCAN-based simulations
and is consistent with previous experimental observations
of aqueous Zn halide solutions.\cite{Dreier1986ZnEXAFS,Dhakal2023ZnHalides}
It is also consistent with the concentration-dependent
changes in the Zn--O and Zn--Cl coordination numbers
discussed in the preceding section. However, because
the Fourier transforms are not phase-corrected and the
Zn--O and Zn--Cl contributions overlap, the apparent
peak positions cannot be directly interpreted as
interatomic distances or coordination numbers.

A particularly interesting result is obtained for the
mixed solution containing 0.25~m \ce{ZnCl2} and 10~m
LiCl. Despite its relatively low \ce{ZnCl2} concentration,
the spectrum exhibits a pronounced contribution in the
Zn--Cl-associated region, highlighting the importance
of chloride availability, rather than \ce{ZnCl2}
concentration alone, in determining the local Zn
environment. The SCAN-based EXAFS spectra obtained
from configurations sampled during enhanced-sampling
MD calculations of the Zn--Cl coordination-number
potential of mean force (PMF) reproduce the main
experimental features, including the distinct
$k$-space oscillations and Fourier-transform profile.
In these calculations, the Zn--Cl coordination number
served as the collective variable. The agreement with
experiment indicates that the PMF simulations sample
local coordination environments consistent with the
measured spectra. Therefore, these comparisons
suggest that SCAN-based DPMD captures the principal
changes in local Zn--O and Zn--Cl coordination across
different electrolyte compositions. 
Complementary EXAFS and structural comparisons for the limiting coordination environments are provided in the SM (Figs. S1 and S3).
Nevertheless, the EXAFS spectra do not
uniquely distinguish isolated Zn complexes from chloride-bridged multinuclear structures. In SM (Figure S5), we briefly examine the contribution of Zn--Zn scattering to the EXAFS
spectra using a representative edge-sharing
\ce{Zn2Cl4} dimer. A more comprehensive investigation
of multinuclear structures and their EXAFS signatures
is left for future work.

\subsection{EXAFS: Sensitivity to the Exchange--Correlation Functional and Dispersion Treatment}
\label{sec:functional-dependence}

\begin{figure}[!htbp]
    \centering

    \includegraphics[width=0.95\columnwidth]
    {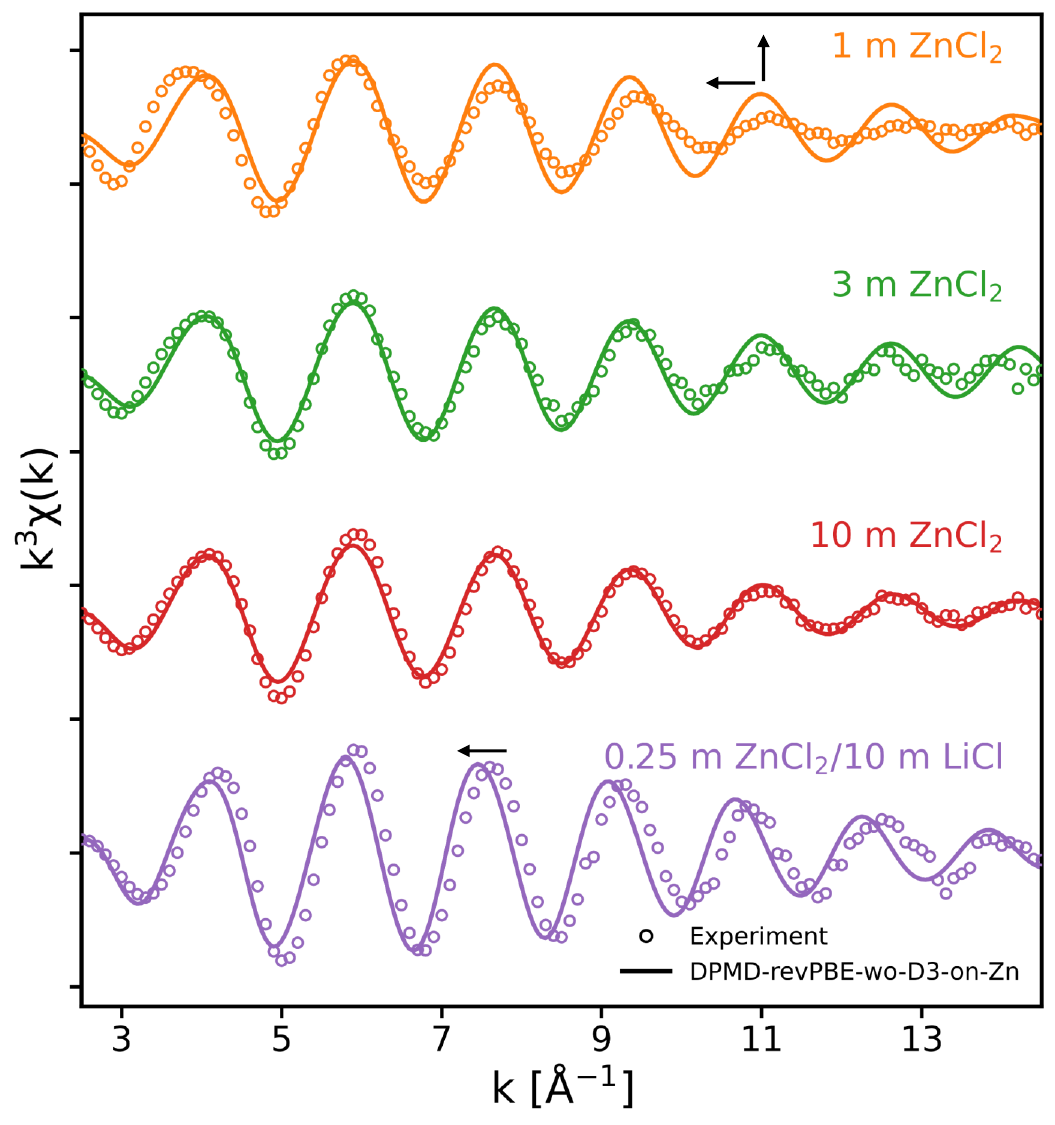}

    \vspace{0.2cm}

    \includegraphics[width=0.95\columnwidth]
    {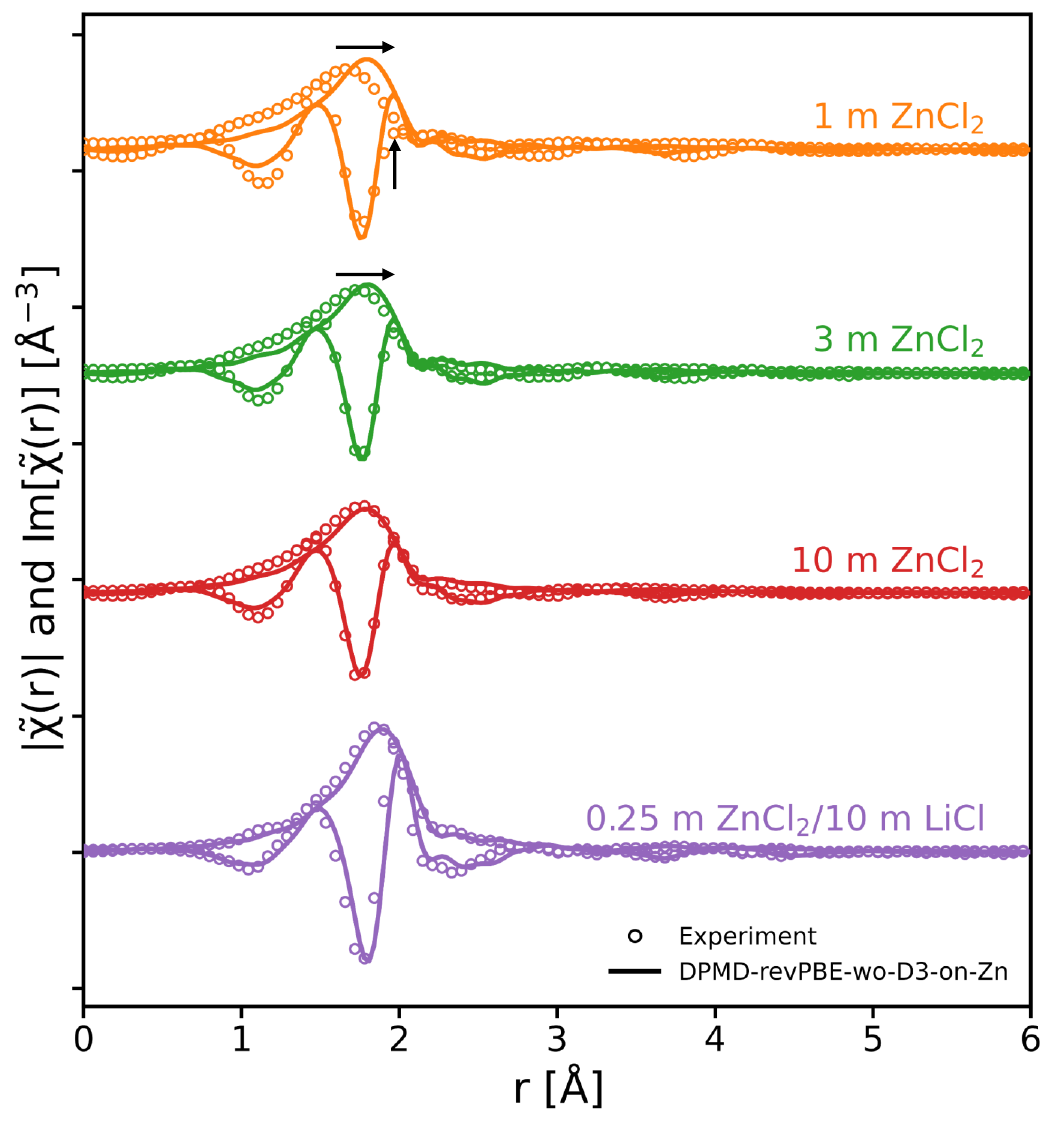}

    \caption{Comparison of experimental (open circles)
    and MD-averaged (solid lines) Zn K-edge EXAFS spectra
    calculated from revPBE-wo-D3-on-Zn trajectories for
    aqueous \ce{ZnCl2} solutions at different concentrations.
    The upper panel shows $k^3\chi(k)$, while the lower
    panel shows the magnitude and imaginary component
    of the Fourier transform. Arrows highlight selected
    differences between calculated and experimental spectra.
    Spectra are vertically offset for clarity.}
    \label{fig:exafs-concentration-revpbe}
\end{figure}

In this section, we examine how differences in the underlying DFT treatment affect the calculated Zn K-edge EXAFS spectra.
By comparing these spectra with our experimental measurements, we assess the effect of excluding D3 dispersion corrections involving Zn on the agreement with experiment.
Figure~\ref{fig:exafs-concentration-revpbe}
compares the experimental spectra with those calculated from the revPBE-wo-D3-on-Zn simulations.
The calculations reproduce several experimental EXAFS features,
particularly at 3 and 10~m \ce{ZnCl2}.
However, the agreement is less satisfactory
at 1~m \ce{ZnCl2} and for the mixed solution containing
0.25~m \ce{ZnCl2} and 10~m LiCl.
Differences in the $k$-space oscillations and the
first-shell Fourier-transform features indicate that
the simulated distributions of local Zn environments
do not consistently reproduce the experimental trends.
Because Zn--O and Zn--Cl scattering contributions overlap
and interfere in the total spectrum, these discrepancies
cannot be assigned uniquely to a particular coordination
number or ligand population.
In contrast, the SCAN-based simulations reproduce the
principal experimental trends across these solution
compositions, as discussed above.
Thus, although removing Zn-related D3 contributions
substantially modifies the revPBE free-energy landscapes,
it does not fully recover the experimentally observed
concentration dependence of the EXAFS spectra
(see Figure~S4 of the SM).

\begin{figure}[tbh]
    \centering
    \includegraphics[width=\columnwidth]{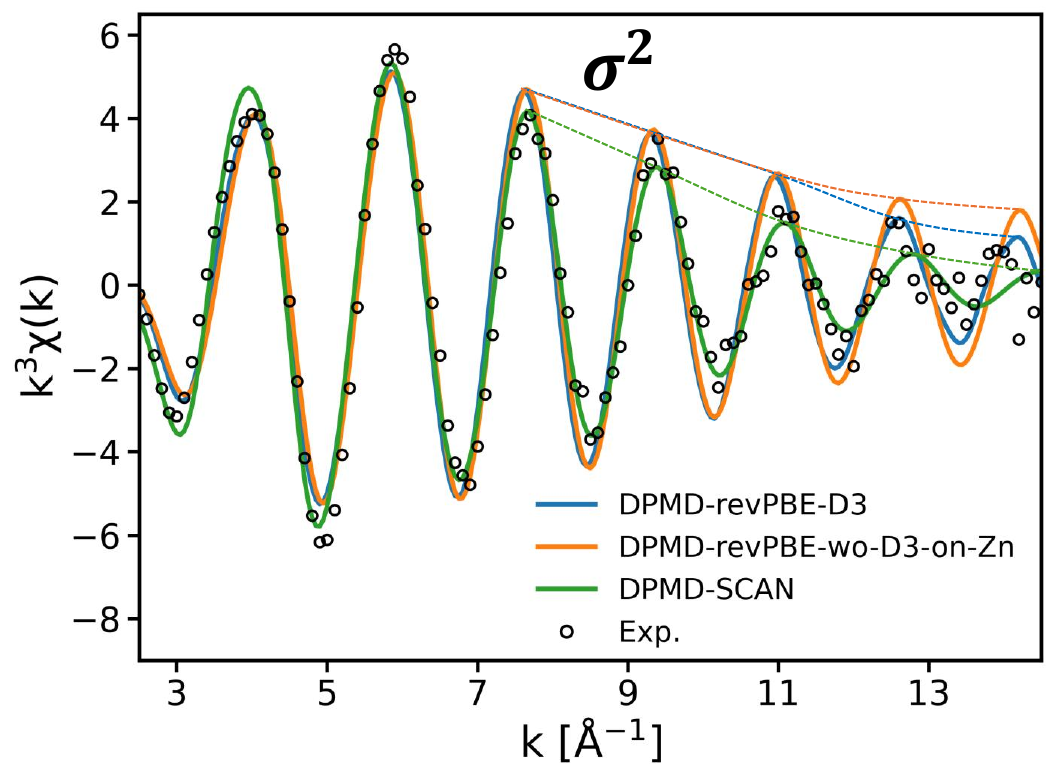}
    \par\vspace{-2mm}
    \includegraphics[width=\columnwidth]{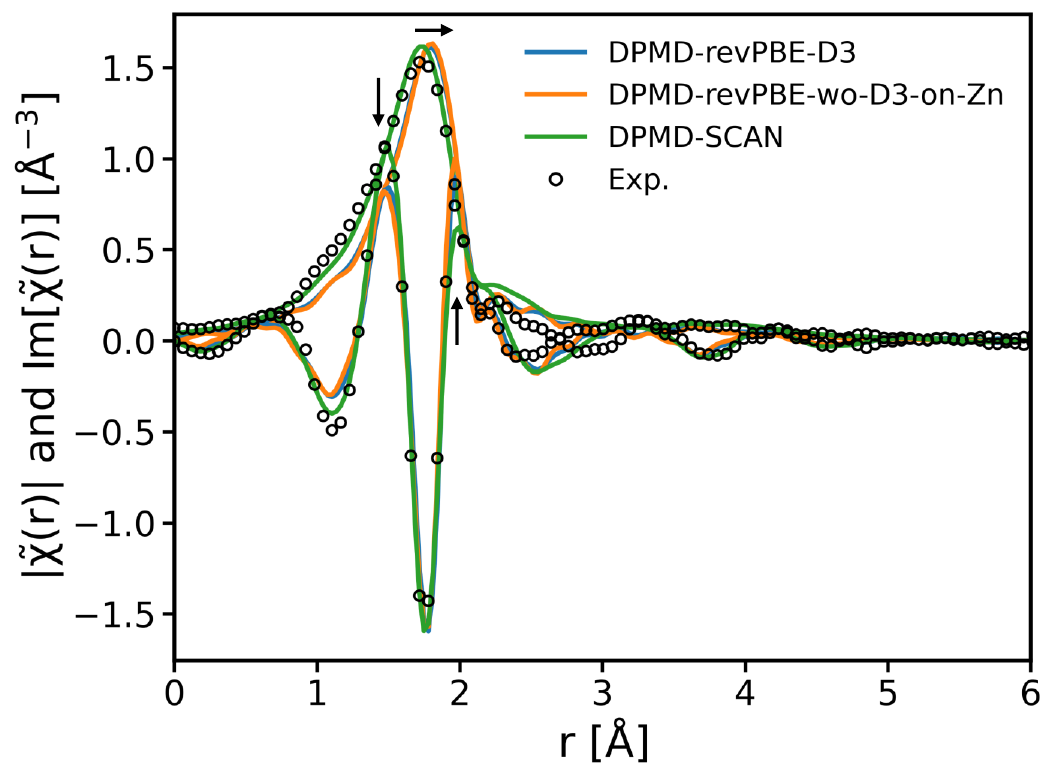}

    \caption{Comparison of Zn K-edge EXAFS spectra
    at 3~m \ce{ZnCl2} obtained using SCAN,
    revPBE-D3, and revPBE-wo-D3-on-Zn.
    (Top) $k$-space EXAFS, $k^3\chi(k)$.
    (Bottom) Magnitude and imaginary component
    of the Fourier-transformed EXAFS.
    Experimental data are shown as open circles.}
    \label{fig:functional-exafs}
\end{figure}

Figure~\ref{fig:functional-exafs} compares the Zn K-edge
EXAFS spectra obtained from all three MLIPs at 3~m \ce{ZnCl2}.
The same FEFF10 procedure was applied to all three
ensembles, allowing us to isolate the influence of
the underlying DFT-trained MLIPs.
Despite substantial differences in Zn--O and Zn--Cl
coordination, all three models reproduce the main
experimental oscillations in $k$ space and predict
similar first-shell Fourier-transform features.
Differences are more apparent at higher $k$, where
the SCAN oscillations are more strongly damped
than those of the two revPBE-based models.
Notably, excluding Zn-related D3 corrections
produces relatively modest changes in the EXAFS
spectra despite altering the local coordination
environment.

The corresponding RDFs and running coordination
numbers reveal substantially different local
environments (Figure~S7 of the SM).
At 3~m, SCAN predicts approximately four water oxygen
neighbors per Zn and fewer chloride neighbors than
either revPBE-based model.
Excluding Zn-related D3 corrections further reduces
Zn--O coordination by approximately 0.4--0.5 relative
to revPBE-D3, with a smaller effect on Zn--Cl coordination.
These results demonstrate that changes in the Zn--Cl
coordination free-energy landscape do not necessarily
produce more SCAN-like hydration structures. Related work on aqueous Na$^+$ and Cl$^-$ found that SCAN produces a more compact and ordered Na$^+$ hydration shell than PBE with Tkatchenko–Scheffler–van der Waals dispersion, attributing this response to changes in charge localization associated with reduced self-interaction error (Li {\em et al.}, Ref~\citenum{Zhaoru-2026}). These findings suggest that the hydration differences observed here may involve exchange–correlation effects on ion–water interactions in addition to the explicit treatment of dispersion.

To understand this spectral similarity, we consider
how the EXAFS signal depends not only on the number
of neighbors but also on their distance distributions
and interference among scattering paths.
For a single-scattering contribution $j$, the
conventional EXAFS expression is

\begin{equation}
\begin{aligned}
\chi_j(k) ={}&
\frac{S_0^2 N_j f_j(k)}{kR_j^2}
\exp\left[-\frac{2R_j}{\lambda(k)}\right]
\exp\left(-2k^2\sigma_j^2\right)
\\
&\times
\sin\left[2kR_j+\delta_j(k)\right],
\end{aligned}
\label{eq:exafs-single-scattering}
\end{equation}

where $N_j$, $R_j$, and $\sigma_j^2$ denote the
coordination number or path degeneracy, mean distance,
and mean-square relative displacement, respectively.
Here, $f_j(k)$ and $\delta_j(k)$ are the effective
scattering amplitude and phase shift, $\lambda(k)$
is the photoelectron mean free path, and $S_0^2$
is the passive-electron amplitude-reduction factor
\cite{Rehr2000XAFS}.

Unlike running coordination numbers, which integrate
the RDFs to count neighbors within a given distance,
the Fourier-transformed EXAFS reflects scattering
amplitudes, phases, and structural disorder.
Scattering phase shifts cause apparent Fourier-transform peak positions to differ from actual interatomic distances, but also carry element-specific information that helps distinguish coordinating ligands. In mixed Zn--O and Zn--Cl environments, overlapping contributions can interfere constructively or destructively. When their total phases differ by approximately $\pi$ over a substantial $k$ range, destructive interference can obscure individual contributions and hinder accurate determination of their relative coordination numbers.\cite{Dreier1986ZnEXAFS,Cao2026ZnCl2XAS} Analogous difficulties have been demonstrated for mixed Zn--N and Zn--S coordination, where nearly out-of-phase oscillations complicate quantification of the ligand populations.\cite{ClarkBaldwin1998} Thus, although phase information is valuable for ligand identification, neither the magnitude nor the imaginary component of the Fourier transform maps directly onto RDFs or coordination numbers, and similar spectra do not necessarily imply identical speciation.

\begin{figure}[tbh]
    \centering
    \includegraphics[width=\columnwidth]{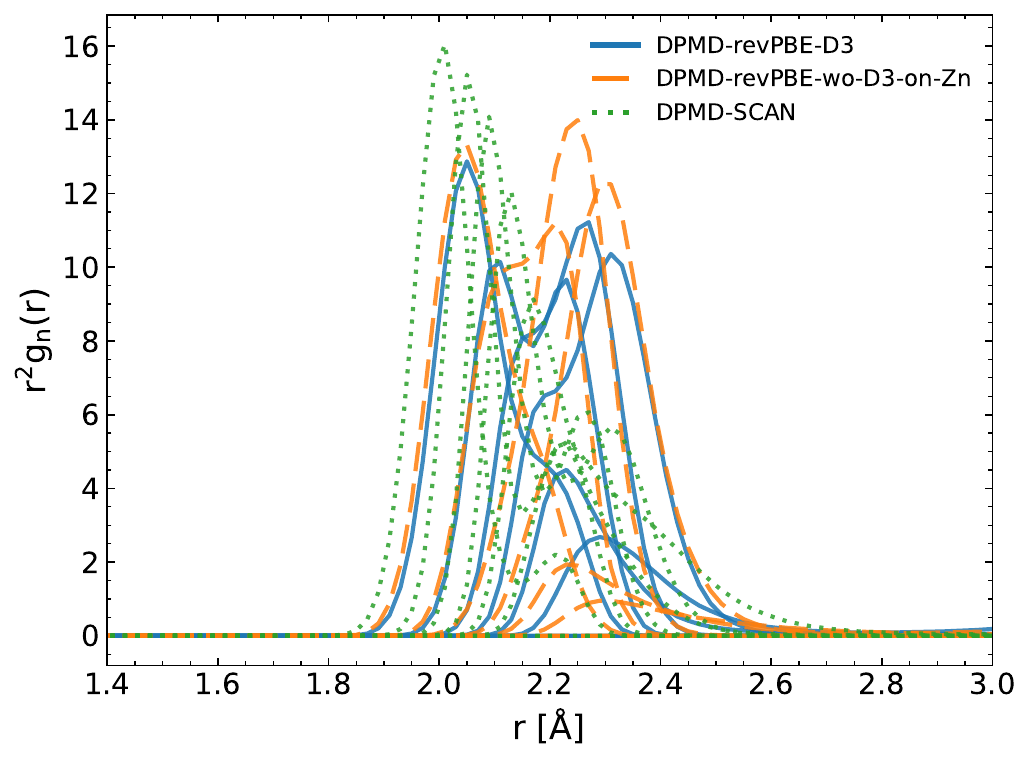}
    \par\vspace{-2mm}
    \includegraphics[width=\columnwidth]{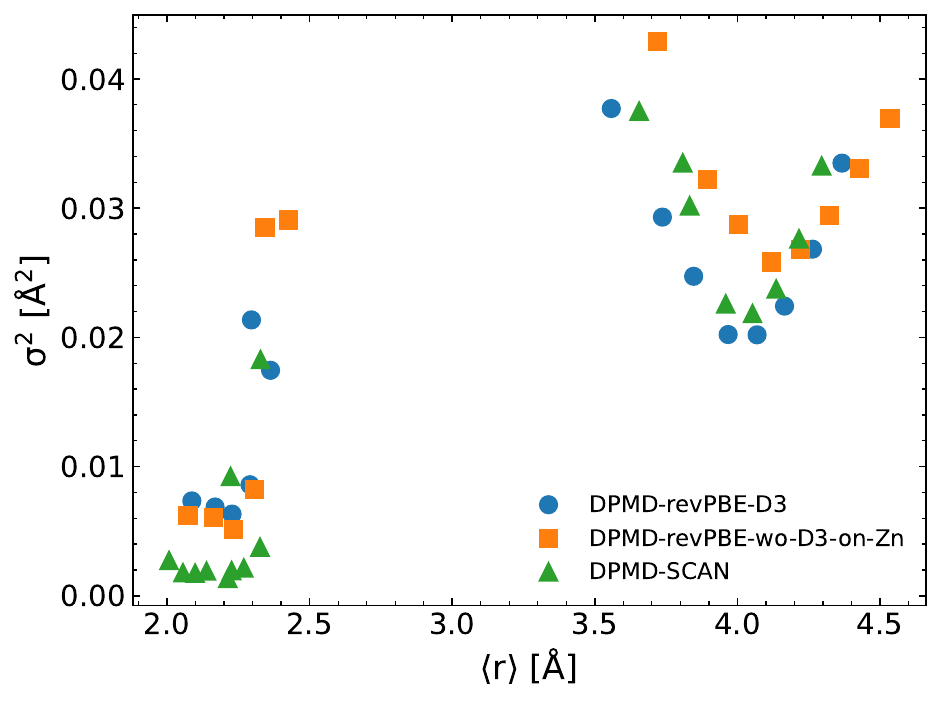}

    \caption{Neighbor-resolved analysis of Zn--X
    ($\mathrm{X}=\mathrm{O},\mathrm{Cl}$)
    coordination at 3~m \ce{ZnCl2}.
    (Top) Incremental radial distributions,
    $r^2g_n(r)$, for distance-ranked Zn--X neighbors.
    (Bottom) Corresponding effective radial variances,
    $\sigma_n^2$, plotted against their mean distances,
    $\langle r_n\rangle$. These variances characterize
    the structural disorder of individual neighbor-rank
    distributions rather than independently fitted EXAFS
    Debye--Waller factors.}
    \label{fig:functional-dwf}
\end{figure}

To quantify the role of structural disorder, we
analyzed individual Zn--X ($\mathrm{X}=\mathrm{O},\mathrm{Cl}$)
neighbors ranked by distance from the central Zn atom,
following Ref.~\citen{duignan2020quantifying}.
Denoting the distance of the $n$th-nearest neighbor
by $r_n$, the normalized radial probability density
obtained from its incremental RDF $g_n(r)$ is

\begin{equation}
P_n(r) =
\frac{r^2g_n(r)}
{\displaystyle\int_{r_{\min}}^{r_{\max}}
r^2g_n(r)\,dr},
\label{eq:incremental-probability}
\end{equation}

where conventional spherical-shell normalization
is assumed. The corresponding mean distance and
radial variance are

\begin{equation}
\langle r_n\rangle =
\frac{
\displaystyle\int_{r_{\min}}^{r_{\max}}
r^3g_n(r)\,dr
}{
\displaystyle\int_{r_{\min}}^{r_{\max}}
r^2g_n(r)\,dr
},
\label{eq:incremental-mean}
\end{equation}

and

\begin{equation}
\sigma_n^2 =
\frac{
\displaystyle\int_{r_{\min}}^{r_{\max}}
(r-\langle r_n\rangle)^2r^2g_n(r)\,dr
}{
\displaystyle\int_{r_{\min}}^{r_{\max}}
r^2g_n(r)\,dr
},
\label{eq:incremental-dwf}
\end{equation}

respectively. When a neighbor-rank distribution is
multimodal, the integration limits can be restricted
to individual peaks (Figure~S8 of the SM). The resulting
moments therefore characterize the selected portion
of the distribution. Figure~\ref{fig:functional-dwf} shows the incremental
radial distributions and corresponding radial variances for the three MLIPs.

The collective radial variance of a coordination shell
includes both fluctuations of individual neighbors and
differences among their mean distances.
For a given scatterer species, with normalized
shell-occupancy weights $w_n$, mean distances
$\mu_n=\langle r_n\rangle$, individual variances
$\sigma_n^2$, and $\bar r=\sum_n w_n\mu_n$,
this variance can be written as

\begin{equation}
\sigma^2 =
\sum_n w_n
\left[
\sigma_n^2+(\mu_n-\bar r)^2
\right],
\qquad \sum_n w_n=1.
\label{eq:collective-dwf}
\end{equation}

The first term describes fluctuations within individual
neighbor populations, while the second accounts for
differences in their mean distances.
For equivalent single-scattering contributions with
an approximately Gaussian distance distribution,
this variance corresponds to the second cumulant
governing the EXAFS Debye--Waller damping,
$\exp(-2k^2\sigma^2)$.
This relationship provides a possible structural
explanation for the differences in high-$k$ amplitudes
observed in the top panel of
Figure~\ref{fig:functional-exafs}.
However, because the total EXAFS signal also involves
interference among scattering paths, the collective
radial variance cannot generally be equated with a
single effective EXAFS Debye--Waller parameter.

Figure~\ref{fig:functional-dwf} shows that SCAN
generally produces narrower distance distributions
for the closest Zn--X neighbors than either revPBE
model, consistent with its stronger first-shell
EXAFS amplitude. Both revPBE models exhibit larger
variances for several short-distance neighbors,
while revPBE-wo-D3-on-Zn shows particularly broad
distributions near the first-shell boundary.
At larger distances (3.5--4.5~\AA), the variances
increase substantially for all three models,
indicating greater structural disorder.

The importance of this disorder is evident from
the EXAFS damping factor $\exp(-2k^2\sigma^2)$.
At $k=10$~\AA$^{-1}$, increasing $\sigma^2$
from 0.006 to 0.020~\AA$^2$ reduces this factor
from approximately 0.30 to 0.02.
Together with distance-dependent attenuation,
this explains why more distant, disordered
neighbors contribute weakly to EXAFS
despite increasing the running coordination numbers.
It is worth stressing that in the EXAFS calculations,
these effects are captured through ensemble averaging
of the configuration-dependent FEFF spectra, while
the neighbor-resolved variances provide a plausible
structural interpretation of the observed damping.

Overall, these results demonstrate that agreement
with experimental EXAFS does not necessarily imply
agreement in the underlying Zn coordination structure.
Distance disorder and interference among scattering
paths can allow distinct structural ensembles to
produce similar spectra. Consequently, EXAFS alone
cannot uniquely resolve Zn speciation, particularly
the distinction between isolated and chloride-bridged
complexes. Benchmarking across a wide concentration
range, combined with structural and free-energy
analyses, therefore provides a more stringent assessment
of the underlying models than agreement at a single
composition.

\subsection{Classical MD versus Universal and Custom MLIPs}
\label{sec:classic-universal-custom}

Beyond agreement with EXAFS, the accuracy of the
underlying interaction model and adequate sampling
of Zn--Cl coordination environments are important
considerations. As discussed in
Sec.~\ref{sec:speciation_convergence}, CMD speciation
populations converge more slowly than those from
DPMD-SCAN: independent CMD trajectories retain
noticeable differences after 50~ns, whereas the
dominant DeepMD species populations become largely
independent of their initial configurations within
25~ns. Although the CMD speciation populations are not fully converged, we compare the corresponding EXAFS spectra with those obtained from MLIP-MD as an approximate assessment of the CMD predictions, acknowledging the remaining sampling uncertainty.

\begin{figure}[!t]
    \centering
    \includegraphics[width=\columnwidth]{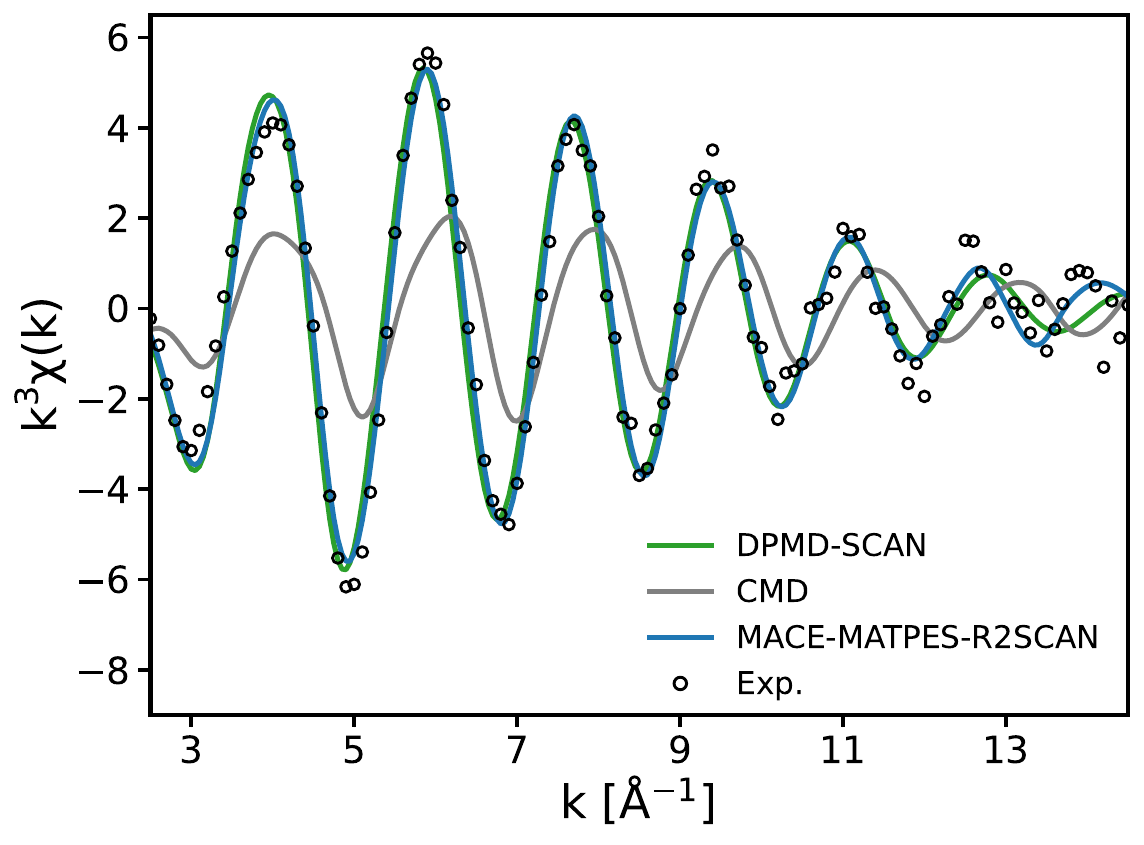}
    \par\vspace{-2mm}
    \includegraphics[width=\columnwidth]{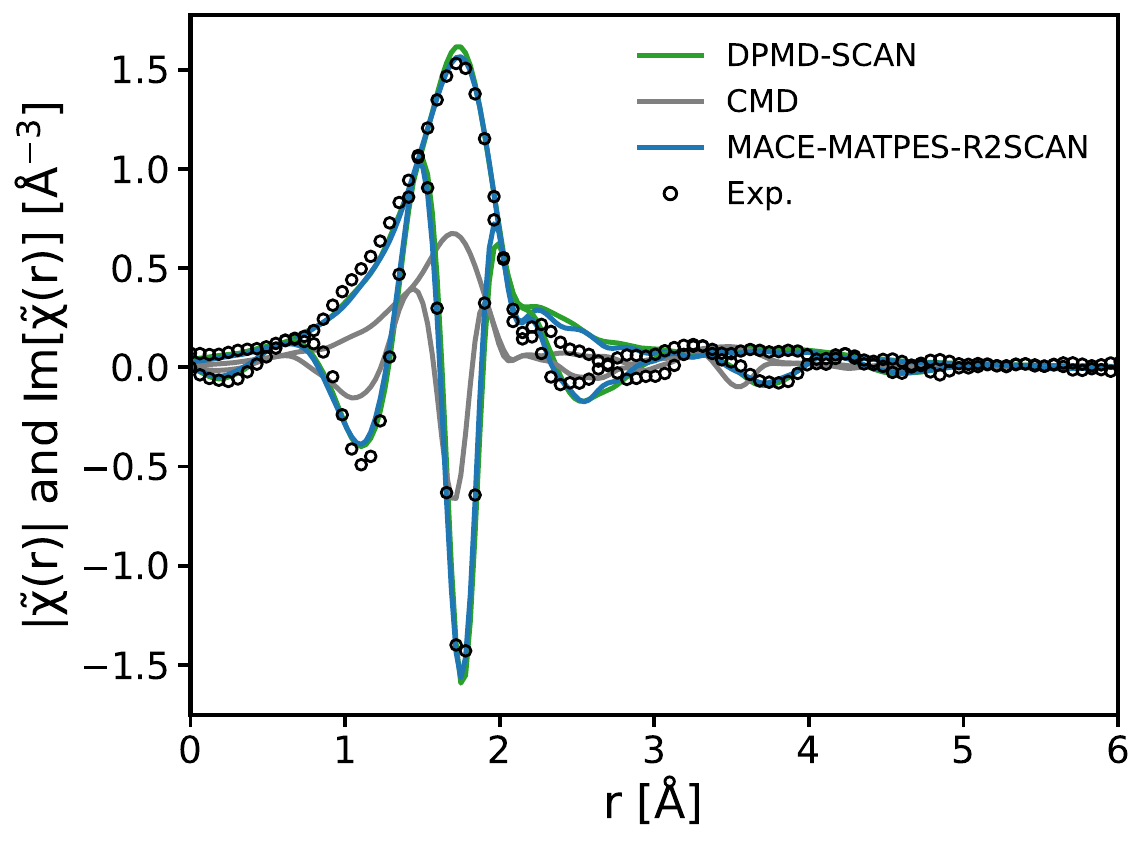}

    \caption{Comparison of experimental Zn K-edge EXAFS
    spectra at 3~m \ce{ZnCl2} with predictions from CMD,
    the universal MACE-MATPES-r2SCAN potential, and
    the custom DPMD-SCAN potential.
    (Top) $k$-space EXAFS, $k^3\chi(k)$.
    (Bottom) Magnitude and imaginary component of
    the Fourier-transformed EXAFS.
    DPMD-SCAN and MACE-MATPES-r2SCAN yield similar
    EXAFS spectra, while CMD shows more pronounced
    deviations from experiment.}
    \label{fig:cmd-exafs}
\end{figure}

\begin{figure}[!t]
    \centering
    \includegraphics[width=\columnwidth]{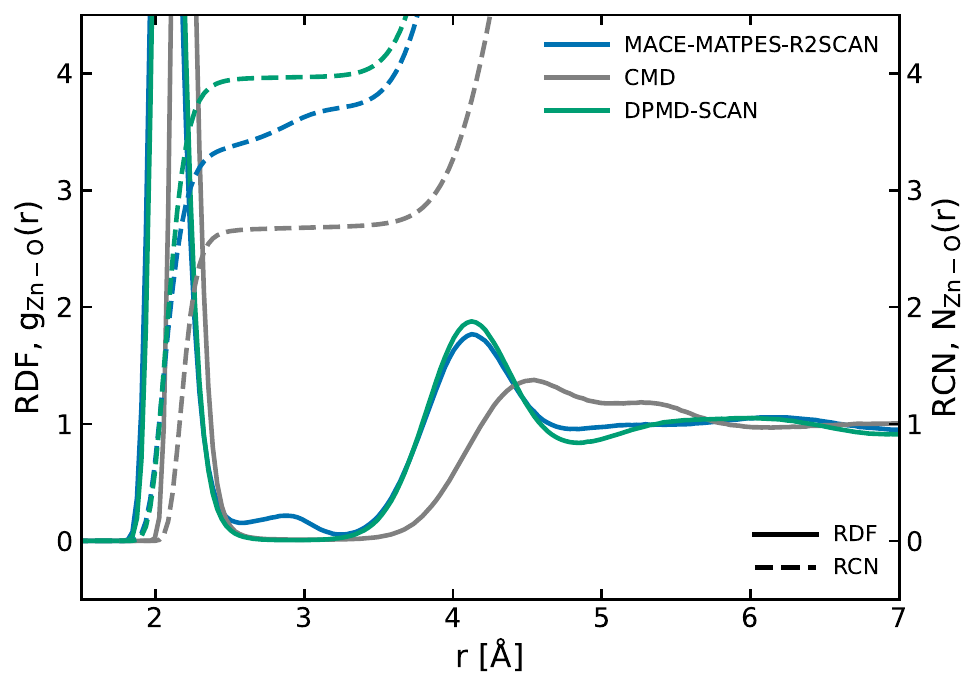}
    \par\vspace{-2mm}
    \includegraphics[width=\columnwidth]{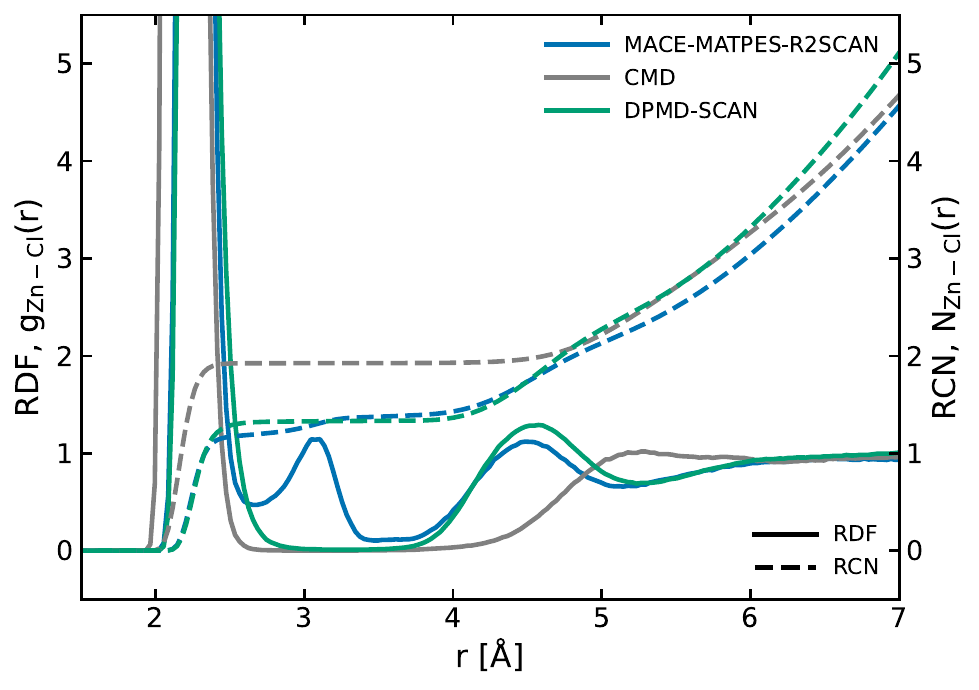}

    \caption{Comparison of local coordination structures
    at 3~m \ce{ZnCl2} obtained using CMD,
    the universal MACE-MATPES-r2SCAN potential,
    and the custom DPMD-SCAN potential.
    Radial distribution functions (RDFs, solid lines)
    and running coordination numbers (RCNs, dashed lines)
    are shown for (top) Zn--O and (bottom) Zn--Cl pairs.}
    \label{fig:cmd-rdf-rcn}
\end{figure}

\begin{figure}[tbh]
    \centering
    \includegraphics[width=\columnwidth]{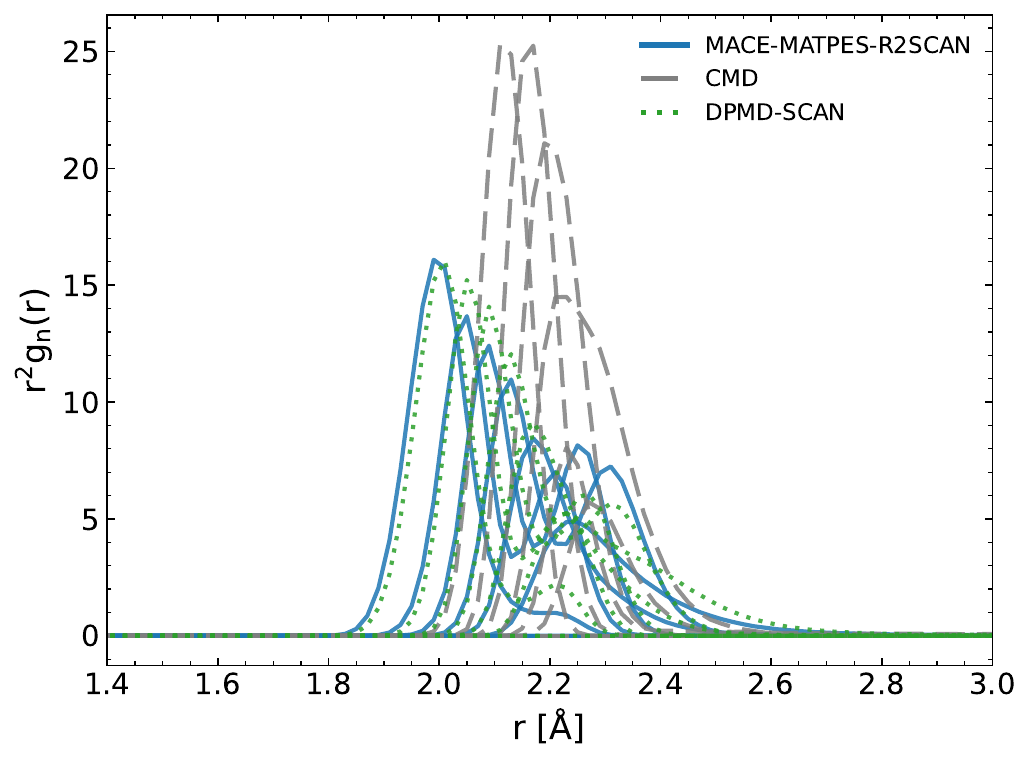}
    \par\vspace{-2mm}
    \includegraphics[width=\columnwidth]{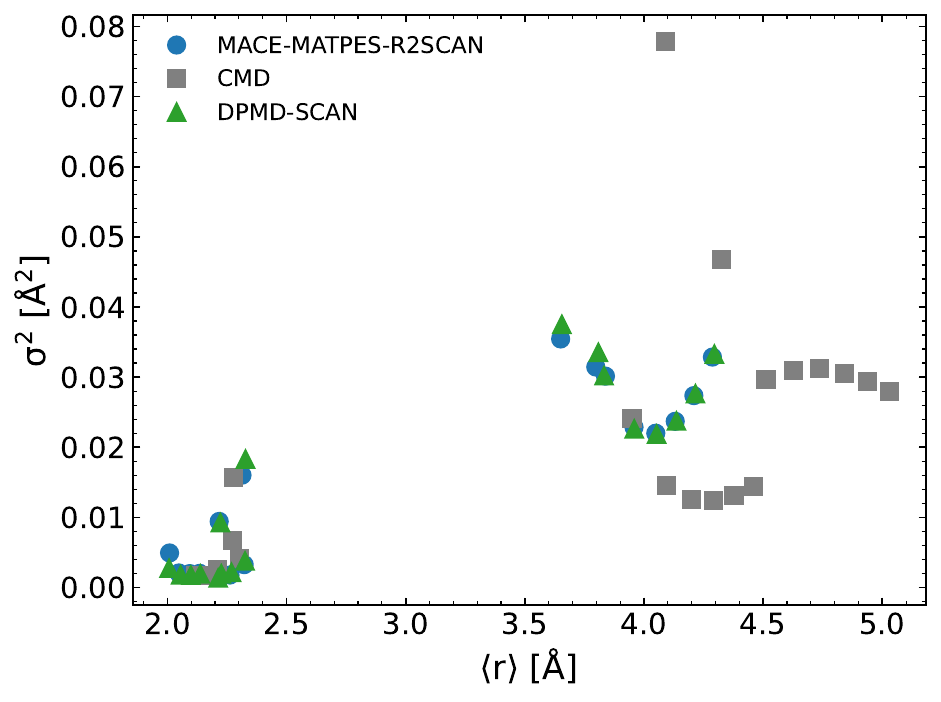}

    \caption{Neighbor-resolved structural analysis of
    Zn--X ($\mathrm{X}=\mathrm{O},\mathrm{Cl}$) at
    3~m \ce{ZnCl2} using CMD, MACE-MATPES-r2SCAN,
    and DPMD-SCAN.
    (Top) Incremental radial particle distribution
    functions (Inc-RPDFs) and (bottom) corresponding
    effective radial variances for distance-ranked
    Zn--X neighbors. Each variance is plotted against
    the mean distance of the corresponding neighbor.
    DPMD-SCAN and MACE-MATPES-r2SCAN predict
    broadly similar distance distributions and
    radial variances, whereas CMD yields more
    localized first-shell environments.}
    \label{fig:cmd-dwf}
\end{figure}

Figure~\ref{fig:cmd-exafs} compares the EXAFS spectra
obtained from the three simulation approaches at 3~m.
The same FEFF workflow was used for all three
ensembles, so differences among the calculated spectra
reflect the structures sampled by the simulations.
In $k$ space, both DPMD-SCAN and MACE-MATPES-r2SCAN
reproduce the principal experimental oscillations
reasonably well in phase and amplitude.
In contrast, CMD substantially underestimates the
oscillation amplitude and exhibits differences in
phase and oscillation envelope.
Similarly, in Fourier-transform space, both MLIPs
reproduce the dominant first-shell feature more
closely than CMD, which predicts a considerably
weaker peak.

The corresponding Zn--O and Zn--Cl RDFs and running
coordination numbers in Figure~\ref{fig:cmd-rdf-rcn}
reveal substantial differences in local coordination.
CMD predicts fewer first-shell Zn--O neighbors
(approximately 2.7 versus 3.5--4.0 for the MLIPs)
and more Zn--Cl neighbors (approximately 1.9 versus 1.3).
In contrast, DPMD-SCAN and MACE-MATPES-r2SCAN
predict broadly similar first-shell coordination,
although differences remain in their RDFs.
Thus, CMD favors more chloride-rich Zn coordination
environments than either MLIP.

To examine the role of structural disorder,
Figure~\ref{fig:cmd-dwf} compares the incremental
radial distributions and neighbor-resolved effective
radial variances.
DPMD-SCAN and MACE-MATPES-r2SCAN produce broadly
similar distributions, whereas CMD yields narrower
and taller distributions for several first-shell
Zn--X neighbors, with generally smaller radial variances.
Interestingly, this trend is opposite to what might
be expected from the weaker CMD EXAFS amplitude.
For fixed coordination numbers, scattering paths,
and mean distances, smaller radial variances reduce
the Gaussian Debye--Waller damping in
Eq.~\ref{eq:exafs-single-scattering} and tend to
increase the EXAFS amplitude.

Thus, radial disorder alone cannot explain the
weaker CMD EXAFS signal.
Differences in Zn--O and Zn--Cl populations,
path lengths, scattering amplitudes, and interference
among contributions with different phases may also
play important roles.
Because distance-ranked distributions do not fully
resolve these scattering contributions, a path-
or species-resolved EXAFS decomposition would be
needed to establish the origin of the CMD amplitude
reduction quantitatively.

Overall, the more localized first-shell distributions
predicted by CMD do not translate into better agreement
with experimental EXAFS. Nevertheless, the force field
developed by Stack and co-workers has shown good agreement
with neutron diffraction and has provided valuable
insights into ion association in aqueous \ce{ZnCl2}.
\cite{Rampal-2021,Dhakal2023ZnHalides}
The similar spectra obtained from DPMD-SCAN and
MACE-MATPES-r2SCAN suggest that the universal potential
captures important features of the local Zn environments
at 3~m. However, as shown in Fig.~S11 of the Supplemental
Material, the two MACE-MATPES-r2SCAN simulations yield
noticeably different speciation populations, indicating
that the available trajectories are insufficient to
establish convergence of the detailed speciation statistics.
For the systems considered here, the custom DeepMD
potentials are approximately an order of magnitude faster
than MACE-MATPES-r2SCAN, enabling longer MD simulations
and more extensive sampling of Zn--Cl coordination
environments. Furthermore, agreement with experimental
EXAFS at a single concentration does not establish
transferability across different electrolyte compositions.

\begin{figure}[b]
    \centering
    \includegraphics[width=\columnwidth]{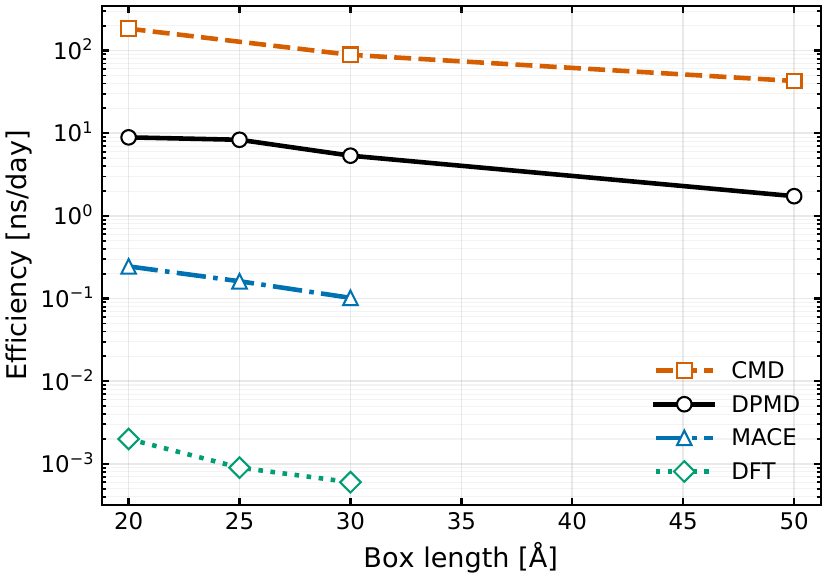}
    \caption{Computational efficiency (ns/day) versus
    simulation box length (20, 25, 30, and 50~\AA{},
    containing 259, 481, 870, and 3848 H$_2$O
    molecules, respectively) for classical MD (CMD),
    custom-trained DeepMD, universally trained MACE,
    and DFT. CMD and DFT used 64 CPU cores, while
    DeepMD and MACE used one NVIDIA H100 GPU.}
    \label{fig:efficiency}
\end{figure}

\subsection{Computational Efficiency}

Figure~\ref{fig:efficiency} reports the computational
efficiency of DFT, DeepMD MLIP, MACE MLIP, and classical
MD simulations used in this work as a function of
simulation box length. The systems considered include
box lengths of 20~\AA{} with 259 H$_2$O molecules,
25~\AA{} with 481 H$_2$O molecules, 30~\AA{} with
870 H$_2$O molecules, and 50~\AA{} with 3848 H$_2$O
molecules. We define computational efficiency as the
physical simulation time generated per day of wall-clock
time (ns/day), calculated from the MD timestep and
wall-clock time per step. CMD and DFT calculations were
performed using 64 CPU cores, whereas DeepMD and MACE
simulations used a single NVIDIA H100 GPU. The
computational efficiencies of MACE and DFT were
estimated from the average time per MD step over
200 steps. CMD is approximately one order of magnitude
faster than DeepMD, DeepMD is approximately two orders
of magnitude faster than MACE, and MACE is approximately
two orders of magnitude faster than DFT. As the box
length increases from 20~\AA{} to 30~\AA{}, the
computational efficiencies of CMD, DeepMD, and MACE
decrease by approximately 50\%, whereas that of DFT
decreases by approximately one-third. Calculations
using MACE and DFT for the 50~\AA{} box could not be
completed because of memory limitations. For a
30~\AA{} box containing 870 H$_2$O molecules, one day
of computation corresponds to approximately 100~ns
of CMD simulation, 5~ns of DeepMD simulation,
100~ps of MACE simulation, and 0.6~ps of DFT simulation.

\begin{figure}[b]
    \centering

    \subfloat[Octahedral Zn coordination]{%
        \includegraphics[width=\columnwidth]
        {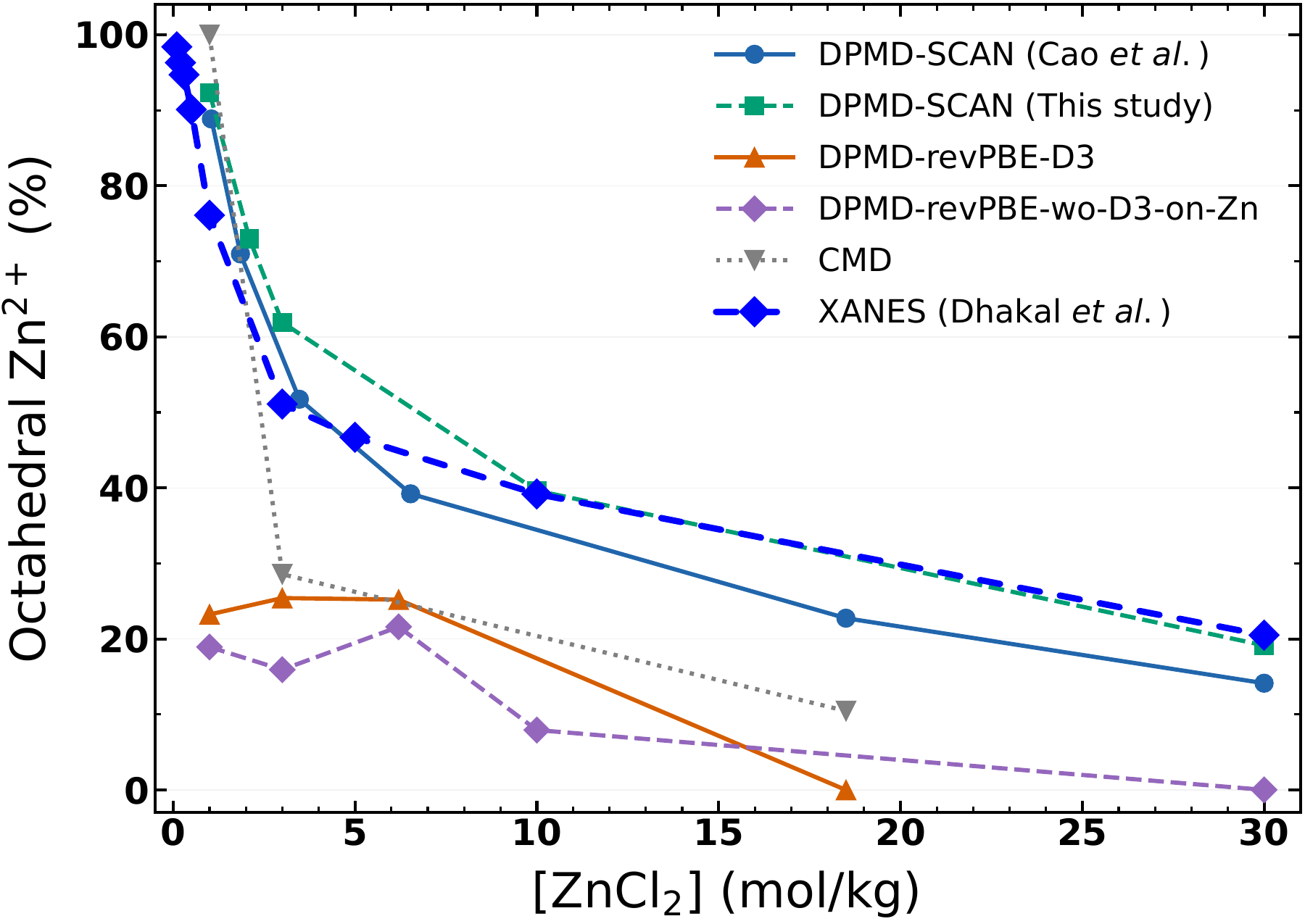}
        \label{fig:octahedral-fraction}
    }

    \subfloat[Average Zn--Cl coordination]{%
        \includegraphics[width=\columnwidth]
        {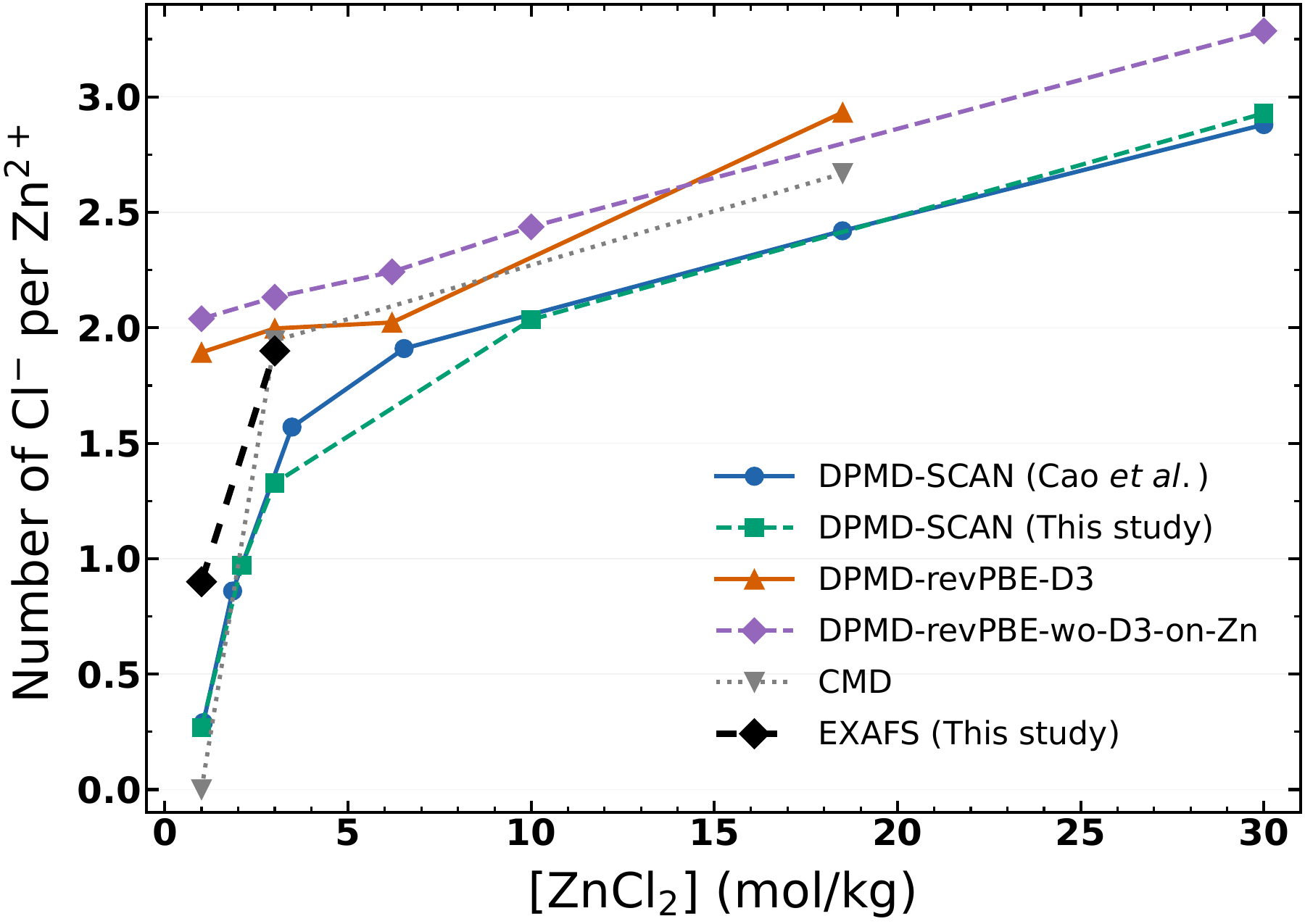}
        \label{fig:average-cl-coordination}
    }

    \caption{Concentration dependence of
    (a) the percentage of Zn ions in octahedral
    coordination environments and
    (b) the average number of Cl neighbors per Zn ion
    in aqueous \ce{ZnCl2} solutions.
    The average Zn--Cl coordination number can exceed
    the bulk Cl/Zn stoichiometric ratio of two because
    chloride ions shared between neighboring Zn centers
    contribute to the coordination shell of each Zn ion.
    At 3~m, the error bars for SCAN, calculated from
    five independent trajectories
    (Figure~\ref{fig:speciation}), are smaller than
    the marker size. Octahedral Zn coordination and the number of Cl$^-$ ions around Zn are evaluated within the first solvation shell, defined using a distance cutoff of 3.0~\AA. }
    \label{fig:Octahedral}
\end{figure}

\subsection{Integrated Assessment of Concentration-Dependent Zn Coordination}

Figure~\ref{fig:Octahedral} summarizes the concentration dependence of two complementary measures of local Zn coordination: the fraction of octahedrally coordinated Zn ions and the average Zn--Cl coordination number. Increasing \ce{ZnCl2} concentration generally decreases the octahedral fraction and increases chloride coordination, consistent with the progressive replacement of water by chloride discussed above.
The DPMD-SCAN results of Cao \textit{et al.} were calculated from Table~S1 of Ref.~\citenum{Cao2026ZnCl2XAS}, and the XANES-derived octahedral fractions were obtained from Figure~5(a) of Ref.~\citenum{Dhakal2023ZnHalides}. The EXAFS coordination numbers were obtained by fitting experimental spectra. 
For the simulations performed in this study, first-shell neighbors were identified using a distance cutoff of $3.0~\text{\AA}$.

The SCAN-trained MLIP broadly reproduces the concentration trends reported by Cao \textit{et al.}. At $30~\mathrm{mol/kg}$, it predicts an octahedral fraction of approximately 19\%, close to the XANES estimate of 21\% and higher than the 14\% reported by Cao \textit{et al.}. Both revPBE-based MLIPs substantially underestimate the octahedral fraction, particularly at low concentrations. The model omitting D3 interactions involving Zn predicts a nearly zero octahedral fraction at $30~\mathrm{mol/kg}$. CMD predicts a higher octahedral fraction than the SCAN and XANES estimates at $1~\mathrm{mol/kg}$, but lower fractions at the higher concentrations examined.

For chloride coordination, both revPBE-based MLIPs predict larger Zn--Cl coordination numbers than DPMD-SCAN, with the largest differences at low concentrations. Omitting D3 interactions involving Zn further increases chloride coordination. At $30~\mathrm{mol/kg}$, this model predicts approximately $3.3$ chloride neighbors per Zn, compared with approximately $2.9$ for both SCAN models. CMD predicts essentially no first-shell chloride at $1~\mathrm{mol/kg}$, but substantial chloride coordination at higher concentrations. 

The SCAN results reach an average Zn--Cl coordination number of approximately 2.0 near $10~\mathrm{mol/kg}$ and approach 3.0 at $30~\mathrm{mol/kg}$. Coordination numbers above two are compatible with the overall \ce{ZnCl2} stoichiometry because a bridging chloride can coordinate multiple Zn centers and contribute to the first-shell count of each. These results support chloride sharing at high concentrations and are consistent with the formation of multinuclear Zn--Cl structures.

\section{Summary \& Conclusions}
\label{conclusion}

This work addresses closely related challenges in concentrated aqueous electrolytes: the transferability of MLIPs across concentrations, the sensitivity of EXAFS to underlying speciation, and the convergence of coordination populations in MD simulations. To investigate these questions, we combine experimental Zn K-edge EXAFS with long-timescale MD simulations of aqueous \ce{ZnCl2}, using two revPBE-based DeepMD potentials developed here and a previously available SCAN-based potential, together with comparisons to a universal MLIP and CMD.

We first benchmarked the EXAFS calculations against two well-defined reference complexes, \ce{[Zn(H2O)6]^2+} and \ce{[ZnCl4]^2-}, and then calculated ensemble-averaged EXAFS spectra from thousands of Zn-centered MD configurations. Among the models examined, DPMD-SCAN provides the most consistent agreement with experimental EXAFS, reproducing the spectral evolution from dilute to concentrated \ce{ZnCl2} solutions. SCAN-based simulations also sample local coordination environments whose calculated spectra capture the main features of mixed \ce{ZnCl2}/LiCl measurements. This performance supports its greater transferability across \ce{ZnCl2} concentrations without retraining at each composition. The MD simulations capture the progressive replacement of water by chloride in the first coordination shell and the increasing importance of chloride-bridged Zn structures at higher concentrations. These trends are consistent with previous SCAN-based simulations and experimental XANES observations.\cite{Cao-2025,Cao2026ZnCl2XAS}
In contrast, the revPBE-D3-based potential reproduces EXAFS at selected concentrations, particularly 3~m, but generally overestimates chloride coordination and provides less consistent agreement across concentrations. Removing the Zn-related D3 contributions does not systematically resolve these discrepancies, highlighting the sensitivity of predicted speciation to the underlying DFT description. 
Importantly, at 3~m, substantially different coordination environments produce similar ensemble-averaged EXAFS spectra, demonstrating that spectral agreement does not uniquely constrain Zn speciation, particularly the populations of isolated and chloride-bridged complexes.

Significant discrepancies with experiment can reveal limitations of the underlying interaction models. At 3~m, CMD shows substantially larger deviations from experimental EXAFS than DPMD-SCAN or MACE-MATPES-r2SCAN, predicting fewer Zn--O and more Zn--Cl neighbors. Despite narrower neighbor-distance distributions, CMD produces weaker EXAFS oscillations, indicating that radial disorder alone cannot explain the spectral differences. However, incomplete equilibration introduces uncertainty in its predicted coordination populations. The universal MACE-MATPES-r2SCAN potential performs considerably better than CMD and produces EXAFS spectra broadly comparable to DPMD-SCAN at 3~m, although this agreement does not establish the accuracy of its speciation predictions or transferability across concentrations. DeepMD is approximately 50 times faster than MACE for a 30~\AA{} simulation box, although CMD remains considerably faster than both MLIPs. Considering its agreement with experiment across \ce{ZnCl2} concentrations and substantially lower computational cost than MACE, DPMD-SCAN offers a favorable balance.

Convergence tests using independent initial configurations demonstrate that the dominant DeepMD coordination populations become largely independent of their initial configurations within 25~ns, whereas CMD retains substantial differences even after longer simulations. 
This slower convergence is consistent with kinetic trapping associated with strong fixed point-charge electrostatic interactions, whereas DeepMD learns electrostatic effects implicitly from DFT reference energies and forces without imposing fixed ionic charges.\cite{Pavel-2013-1,Pavel-2018,Pavel-2019-1,Dinpajooh-2024,Biriukov-2022,Panagiotopoulos-2023,Schran-2024}
Even with DeepMD, less abundant multinuclear aggregates remain more difficult to sample. In future work, we will present related enhanced-sampling approaches to calculate EXAFS spectra across concentrations, reducing the need for long unbiased MD simulations.

The importance of accurate interaction models and adequate sampling becomes especially clear when examining the concentration dependence of Zn coordination and connectivity. As \ce{ZnCl2} concentration increases, octahedral Zn environments become less prevalent, while chloride coordination and bridging increase. DPMD-SCAN better reproduces the available experimental trends: at 30~m, its predicted octahedral fraction (19\%) is close to the XANES estimate (21\%), while average Zn--Cl coordination approaches three, consistent with chloride sharing. In contrast, CMD shows less consistent agreement with the XANES-derived octahedral fractions, overestimating them at 1~m and underestimating them at higher concentrations. The predicted extent of aggregation also remains strongly model-dependent. At 10~m, revPBE-wo-D3-on-Zn predicts approximately 78\% of Zn in aggregates containing three or more Zn centers, compared with less than 1\% for SCAN. These differences underscore why local coordination and molecular connectivity must be analyzed separately.

These contrasting structural predictions also clarify what EXAFS can establish about electrolyte speciation. Our neighbor-resolved analysis shows that distinct coordination ensembles can produce similar EXAFS spectra because of structural disorder, variations in neighbor distances, and interference among scattering paths. Thus, EXAFS provides important constraints on local Zn coordination but does not uniquely determine the populations of isolated and chloride-bridged complexes. The agreement of DPMD-SCAN with EXAFS across concentrations, together with independent XANES observations, strengthens confidence in its description of local structure without uniquely validating its predicted speciation. Reliable speciation predictions therefore require accurate interatomic interactions, converged molecular sampling, ensemble-averaged spectral calculations, and explicit analysis of coordination and connectivity. Overall, these findings establish the importance of assessing MLIP transferability across concentrations and combining rigorous molecular simulations with complementary experimental constraints for predictive speciation in concentrated electrolytes.

\hfill
\section*{Supplementary Material}
The Supplementary Material provides theoretical EXAFS details, complementary EXAFS and structural analyses, and WHAM and coarse-grained simulation consistency checks.

\section*{Acknowledgements}
JC, LDG, and MD were supported by FWP 85666, a U.S. Department of Energy (DOE), Office of Science (SC), Early Career Research Program award in the Basic Energy Sciences, Chemical Sciences, Geosciences, and Biosciences Division, Condensed Phase and Interfacial Molecular Science program, for developing machine-learning interatomic potentials and reduced models of electrolyte solutions, and related analysis workflows.
DMD and MB were supported by the Joint Center for Energy Storage Research, an Energy Innovation Hub funded by the U.S. DOE.  This research used resources of the Advanced Photon Source, an Office of Science User Facility operated for the U.S. DOE Office of Science by Argonne National Laboratory and was supported by the U.S. DOE under contract no. DE-AC02-06CH11357.
NNI, CJM, and GKS were supported by the U.S. DOE, Office of Science, Basic Energy Sciences, Chemical Sciences, Geosciences, and Biosciences Division, Condensed Phase and Interfacial Molecular Science program under FWP 16249; JLF was supported by the same program under FWP 16248.
PNNL is a multi-program national laboratory operated for the U.S. DOE by Battelle Memorial Institute under Contract No. DE-AC05-76RLO 1830. This work was also supported by the resources of the National Energy Research Scientific Computing Center (NERSC), a DOE Office of Science User Facility, using the NERSC Award No. BES-ERCAP0035156 .

\section*{Author Declarations}
\subsection*{Conflict of interest}
The authors have no conflicts to disclose.

\section*{Data Availability}
The data that support the findings of the study are openly available on PNNL DataHub under the DOI \href{https://doi.org/10.25584/3400831}{10.25584/3400831}.
The rest of the data that supports the findings of this study are available within the article and its Supplementary Material. The source code implementations are also available from the corresponding author upon reasonable request.

\bibliography{references}

\end{document}